\documentclass[reprint,superscriptaddress,amsmath,amssymb,aps]{revtex4-2}
\usepackage{natbib}
\usepackage[colorlinks,linkcolor = blue,citecolor=blue,urlcolor=blue,bookmarks=false,hypertexnames=true]{hyperref}
\usepackage{float}
\usepackage{comment}
\usepackage{mathtools}
\usepackage{graphicx}
\usepackage[export]{adjustbox}
\usepackage{dcolumn}
\usepackage{bm}
\usepackage{sidecap}
\usepackage{chemformula}
\usepackage[justification=justified,singlelinecheck=false]{subcaption}
\usepackage{gensymb}

\begin{document}

\title{Spin Wave Dispersion of the van der Waals Antiferromagnet \ch{NiPS3}}

\author{Ritesh Das}
\email{R.D.Das@tudelft.nl}
\affiliation{%
Kavli Institute of Nanoscience, Delft University of Technology, Lorentzweg 1, 2628 CJ, Delft, The
Netherlands
}%

\author{Rob den Teuling}
\affiliation{%
Kavli Institute of Nanoscience, Delft University of Technology, Lorentzweg 1, 2628 CJ, Delft, The
Netherlands
}%

\author{Artem V. Bondarenko}
\affiliation{%
Kavli Institute of Nanoscience, Delft University of Technology, Lorentzweg 1, 2628 CJ, Delft, The
Netherlands
}%

\author{Elena V. Tartakovskaya}
\affiliation{%
Kavli Institute of Nanoscience, Delft University of Technology, Lorentzweg 1, 2628 CJ, Delft, The
Netherlands
}%
\affiliation{%
V.G. Baryakhtar Institute of Magnetism of the NAS of Ukraine,   
36b Vernadsky Boulevard, 03142 Kiev, Ukraine.
}%

\affiliation{%
Institute of Spintronics and Quantum Information, Faculty of Physics and Astronomy, Adam Mickiewicz University, Poznań, Uniwersytetu Poznańskiego 2, 61-614 Poznań, Poland  
}%

\author{Gerrit E. W. Bauer}
\affiliation{%
WPI Advanced Institute for Materials Research, Tohoku University, Sendai, Japan
}%
\affiliation{%
Kavli Institute for Theoretical Sciences, University of the Chinese Academy of Sciences, Beijing, China
}%

\author{Jaime Ferrer}
\affiliation{%
Departamento de Fisica, Universidad de Oviedo, 33007 Oviedo, Spain
}%
\affiliation{%
Centro de Investigacion en Nanomateriales y Nanotecnologia, Universidad de Oviedo-CSIC, 33940, El Entrego, Spain
}%

\author{Yaroslav M. Blanter }
\email{Y.M.Blanter@tudelft.nl}
\affiliation{%
Kavli Institute of Nanoscience, Delft University of Technology, Lorentzweg 1, 2628 CJ, Delft, The
Netherlands
}%

\date{\today}

\begin{abstract}
We calculate the magnon dispersion spectra of the two-dimensional zigzag van der Waals antiferromagnet \ch{NiPS3} for monolayer, bilayer, and bulk systems as a function of an external magnetic field. \textcolor{black}{We calculate the exchange and anisotropy constants in our spin model by first principles.} We can accurately explain the transition from a collinear to a canted ground state for a magnetic field applied normal to the (in-plane) easy-axis, and a spin-flop transition when the field is parallel to it. A topologically protected Dirac nodal line is present and robust with respect to both external and anisotropy fields.
\end{abstract}

\maketitle
\section{Introduction}
The elementary excitations of a magnetic ground state are known as magnons, the quanta of spin waves \cite{magnon,magnonBook,stancilprabhakar, mewaves, comp2}, that are ubiquitous in magnetometry \cite{magnetometer1, magnetometer2}, caloritronics \cite{calori1,calori2,calori3}, and potentially useful applications in computation \cite{comp1,comp2}. They are compatible with complementary metal–oxide–semiconductor (CMOS) technology for logic devices \cite{comp2}, and the information transmitted by spin waves does not suffer from Ohmic losses \cite{comp2}. \par

Promising platforms for studying spin waves are two-dimensional (2D) van der Waals (vdW) magnetic materials \cite{vdw1,vdw2,vdw3,vdw4}. 2D vdW magnets can be exfoliated to monolayers or a few layers \cite{vdw2, bilsynth}. The competing energies in these materials give rise to highly tunable magnon dispersions and spin wave dynamics \cite{modulation, vdw3}. \ch{NiPS3} is a vdW antiferromagnet with a hexagonal zigzag ground state, where the spins are predominantly in-plane along an easy-axis \cite{wildesstrcture, lanconmonolayerscatter, wildes2022magnetic, nips3bulkneutronscatter}, with a small out-of-plane component \cite{wildes2022magnetic, nips3bulkneutronscatter}. \ch{NiPS3} has been predicted to host quantum phenomena such as non-trivial topology \cite{topologyzigzag1}, anomalous scattering \cite{nips3bulkneutronscatter}, and non-trivial magnetic order \cite{bilsynth}. To describe these effects, a clear understanding of the Hamiltonian and spin wave dynamics in \ch{NiPS3} is essential. \par

Spin waves in \ch{NiPS3} have been studied in bulk samples through neutron scattering experiments \cite{lanconmonolayerscatter, nips3bulkneutronscatter, wildes2022magnetic}, which provide dispersions across the entire Brillouin zone. Several models describe the magnon dispersions in monolayer \cite{lanconmonolayerscatter} and bulk \ch{NiPS3} samples \cite{lanconmonolayerscatter, wildes2022magnetic, nips3bulkneutronscatter}, but outstanding questions remain, such as the dependence on externally applied magnetic fields, crystal anisotropy constants, and dipolar interactions. According to the Mermin-Wagner theorem, a Heisenberg magnet cannot exhibit long-range magnetic order in the 2D limit without anisotropy \cite{merminog}. Although all models of \ch{NiPS3} predict easy-axis anisotropy, long-range magnetic order exists only in bilayer samples and vanishes in monolayers \cite{bilsynth}. This suggests that interlayer exchange is crucial for maintaining long-range order. The bilayer model is therefore a minimal model for studying low-dimensional antiferromagnetism in many vdW systems. Recently, observation of the spin resonance in bilayer \ch{NiPS3} \cite{bilayercanting} makes a model for the bilayer particularly relevant. \par

Here, we present magnon dispersion and resonance frequency for monolayer and bilayer \ch{NiPS3} as solutions of the Landau-Lifshitz equation for an in-plane ground state \cite{lanconmonolayerscatter}. \textcolor{black}{We compute the exchange and anisotropy constants from first-principles  that predict an in-plane zigzag antiferromagnetic ground state.} We investigate in detail the parameter dependence of nearly degenerate magnons closer to the Dirac points. We also compare the resonance frequencies as function of the external field with available references \cite{bilayercanting, dima}. The paper is organized as follows: In Sec. II, we elaborate on the model and describe the Hamiltonian governing the ground state and its excitations. We show how to linearize the Landau-Lifshitz equation (LL equation) to compute spin wave dispersion. \textcolor{black}{In Sec. III, we describe the first-principles methods used to compute the exchange and anisotropy constants.} In Sec. IV, we present the magnon dispersion for monolayer  and bilayer \ch{NiPS3}, examining the dependence of the spin wave band structure on external fields and anisotropy constants. In Sec. V, we report the resonance frequencies as a function of applied magnetic field. We conclude in Sec. VI.

\section{Model}
The magnetic structure of \ch{NiPS3} in the crystallographic unit cell, along with the corresponding Brillouin zone, is shown in Fig. 1. \ch{NiPS3} features a hexagonal zigzag antiferromagnetic lattice \cite{lanconmonolayerscatter, nips3bulkneutronscatter, topologyzigzag1, wildes2022magnetic, wildesstrcture}, with magnetic moments localized on the \ch{Ni} atoms. Recent neutron scattering experiments have proposed two similar yet conflicting models for the magnetic structure of \ch{NiPS3}. Ref. \cite{lanconmonolayerscatter} reports a strong easy-axis along a single in-plane direction, resulting in spins that are confined to the plane in the ground state (see Fig. 1(a)). In this model, the spins lie along the $z$-axis. The $x$($y$) axis is perpendicular to the $z$-axis, and corresponds to the direction in (out of) the crystallographic plane. In contrast, Ref. \cite{wildes2022magnetic} finds a weak easy-axis along the $z$-direction, along with a strong out-of-plane hard-axis making an angle of $73.02^\circ$ with the plane. This results in the spins lying slightly out-of-plane in the ground state (see Fig. 1(c)). The model proposed in Ref. \cite{wildes2022magnetic} is more recent and is considered to provide a more accurate description of the magnetic structure. Ref. \cite{nips3bulkneutronscatter} also finds values that agree with Ref. \cite{wildes2022magnetic}.

For simplicity we adopt a model with both an easy and a hard-axis \cite{wildes2022magnetic}, assuming that the hard-axis is perfectly perpendicular to the plane (rather than at the observed angle of $73.02^\circ$). This simplification results in a ground state where the spins lie entirely in the plane. \textcolor{black}{Noting that our first-principles calculations (see Sec. III) also fail to find the observed slightly out-of-plane configuration}. The magnetic primitive unit cell contains four atoms in a rectangular parallelepiped, as depicted in Fig. 1(b) together with the first Brillouin zone. In the following subsections, we specify the Hamiltonian and determine the ground state by minimizing the corresponding free energy. The Landau-Lifshitz (LL) equation is then derived by linearizing the free energy for small excitations around the ground state, from which we calculate the spin-wave dispersion for various system parameters.
\begin{figure}[]
 \begin{center}
 \begin{subfigure}{0.2\textwidth}
    \includegraphics[width=\linewidth]{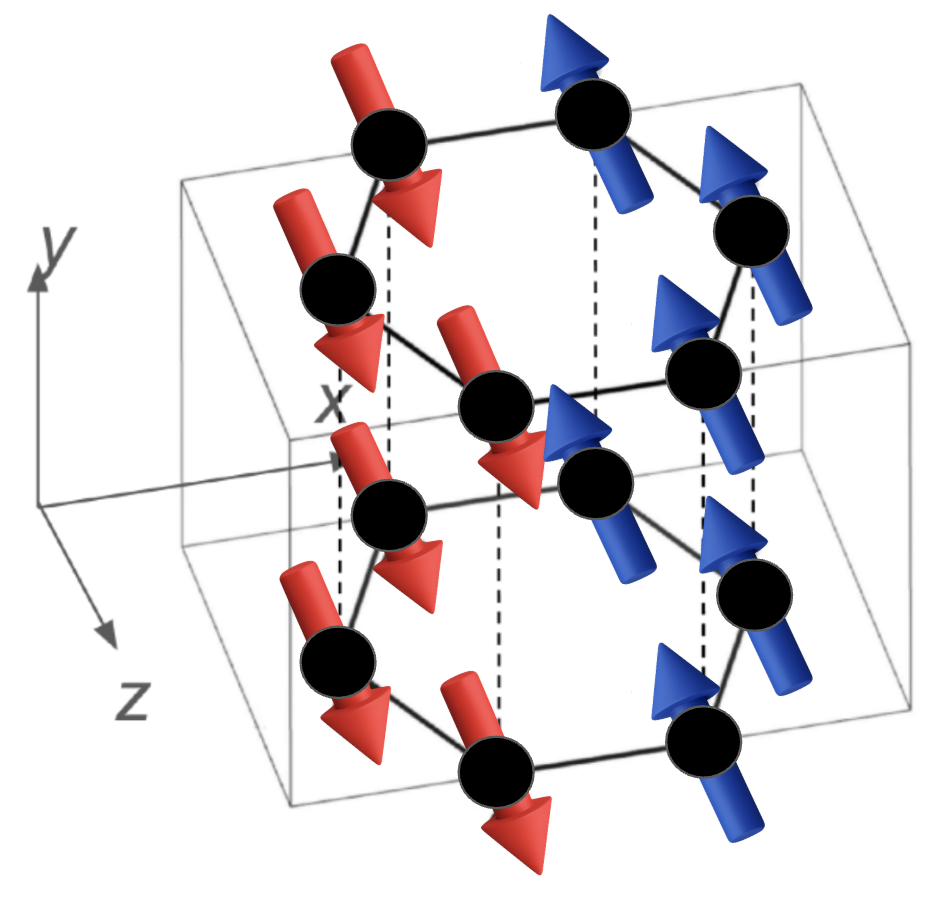}
    \caption{} 
  \end{subfigure}
  \begin{subfigure}{0.25\textwidth}
    \includegraphics[width=\linewidth]{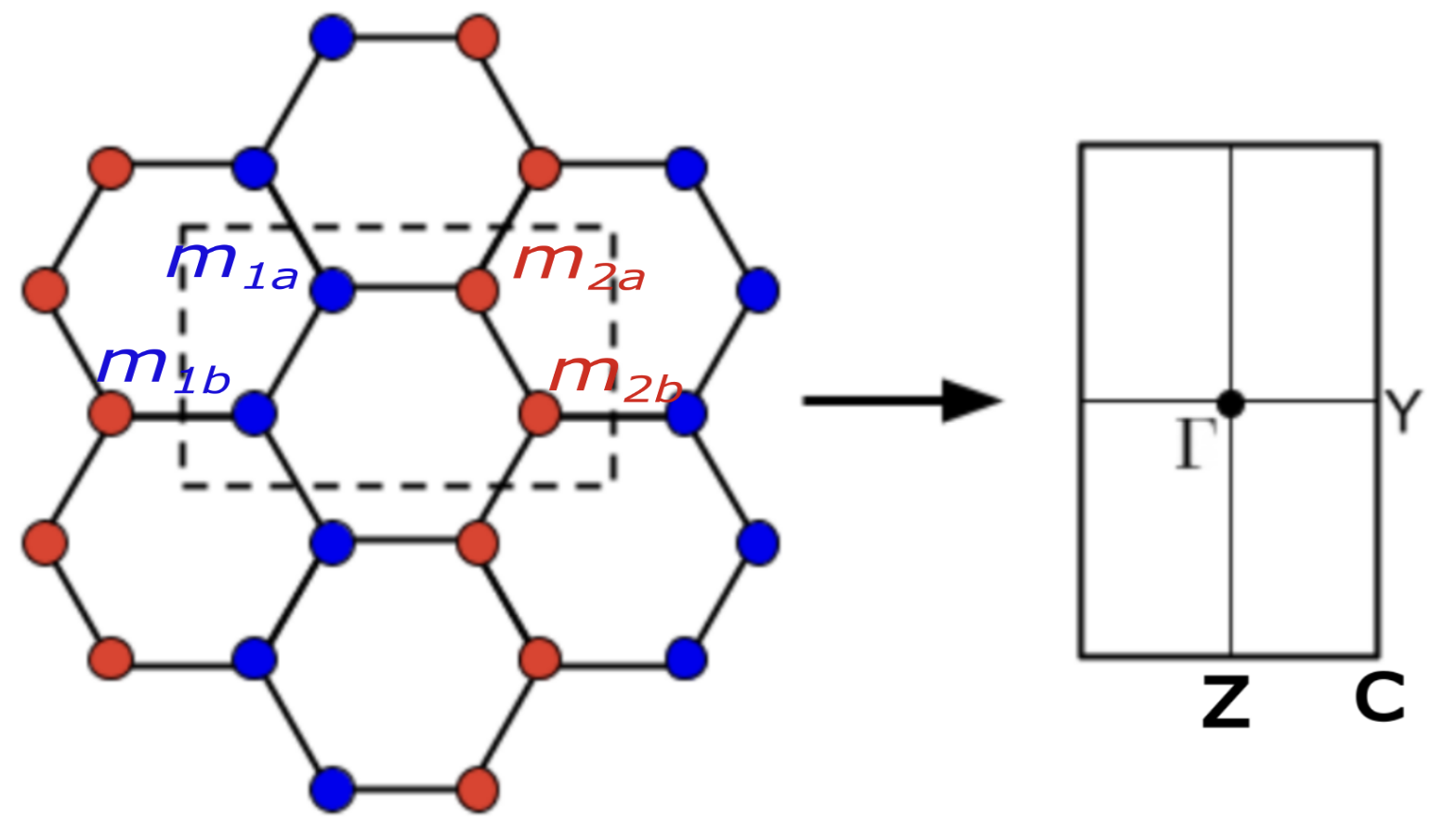}
    \caption{} 
  \end{subfigure}\\
 \begin{subfigure}{0.45\textwidth}
    \includegraphics[width=\linewidth]{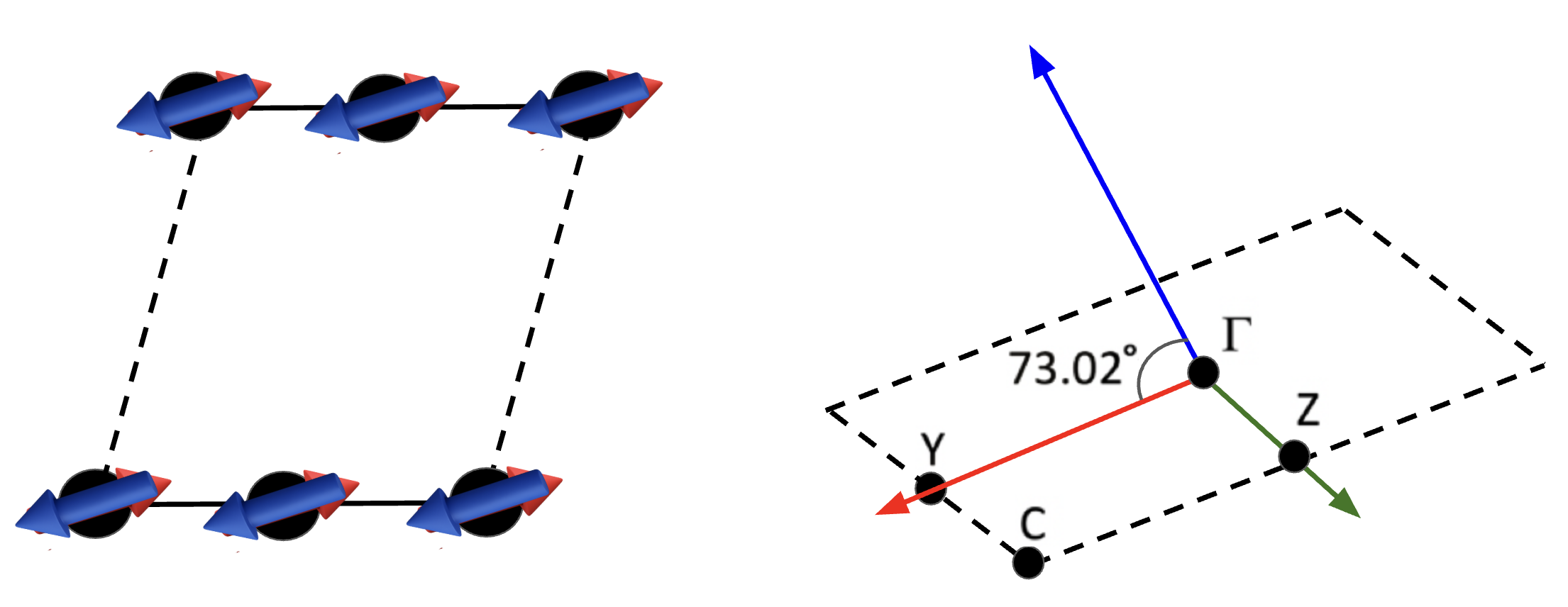}
    \caption{} 
  \end{subfigure}
  \end{center}
 \caption{\centerlast
 (a) Bilayer \ch{NiPS3} assumes a zigzag antiferromagnetic hexagonal lattice with in-plane magnetic moments. Red arrows represent spin-up (along z) and blue arrows represent spin down directions. (b) Unit cell in real space (left) and the corresponding Brillouin zone in reciprocal space (right). The red dots represent spin up and blue dots spin down magnetic moments. (c) illustrates the observed magnetic structure \cite{wildes2022magnetic} with a hard-axis at an angle of $73.02 \degree$ with the plane and a ground state in which spins are slightly pulled out-of-plane.} 
 \end{figure}
\begin{figure}[]
 \begin{center}
    \begin{subfigure}{0.15\textwidth}
    \includegraphics[width=\linewidth]{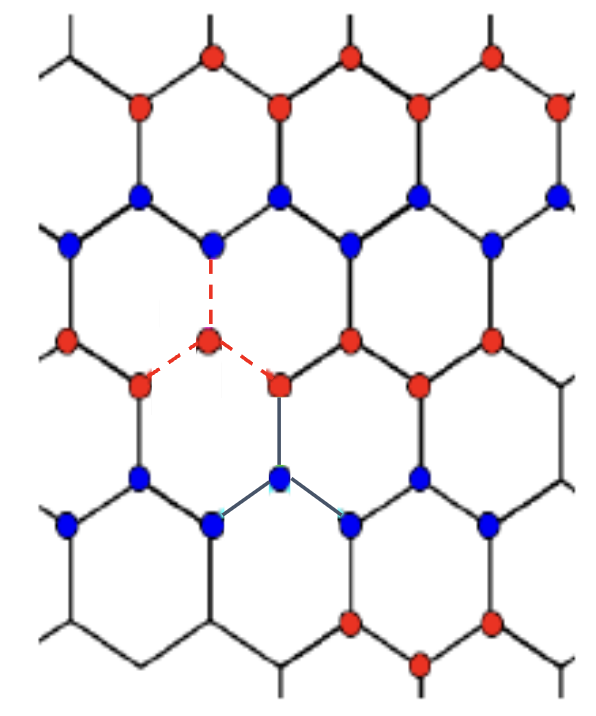}
    \caption{} 
  \end{subfigure}
  \begin{subfigure}{0.15\textwidth}
    \includegraphics[width=\linewidth]{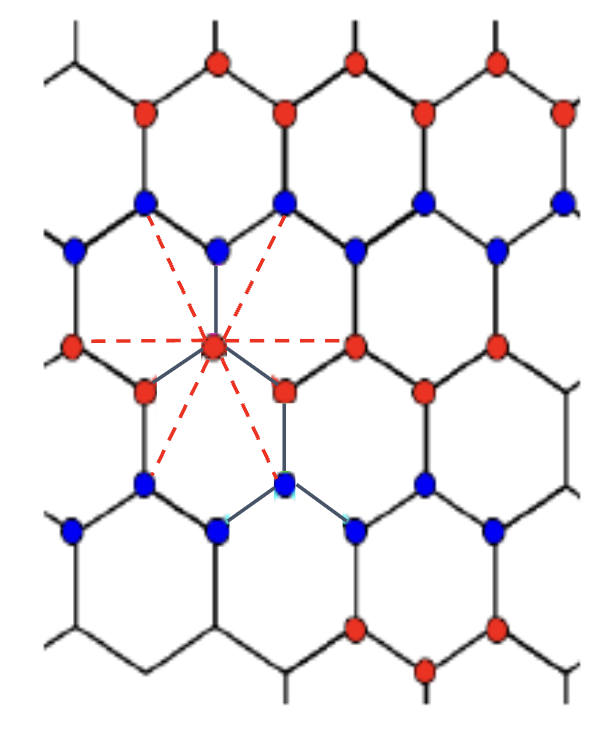}
    \caption{} 
  \end{subfigure}
  \begin{subfigure}{0.15\textwidth}
    \includegraphics[width=\linewidth]{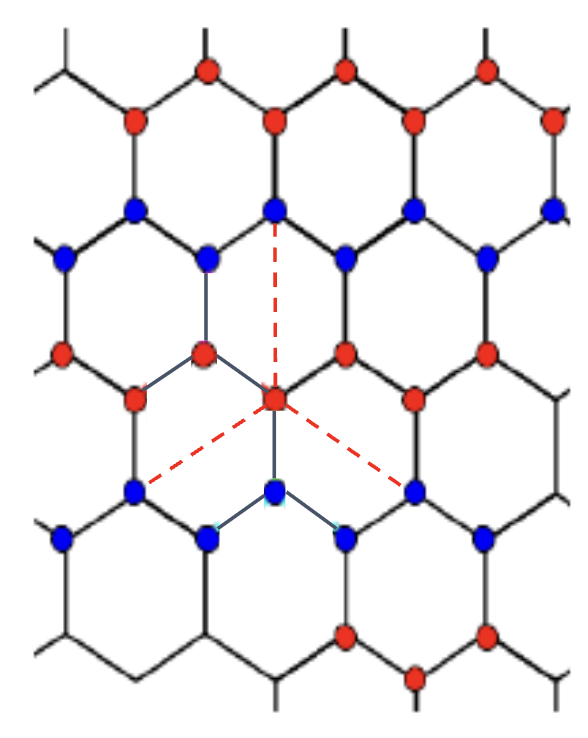}
    \caption{} 
  \end{subfigure}
  \end{center}
 \caption{\centerlast Exchange interactions in \ch{NiPS3}. The red (blue) dots represent spin-up (spin-down) local moments. The black lines denote the bonds in the crystal, the red dashed lines denote the path to the (a) nearest neighbors, (b) next-nearest neighbors, and (c) next-next-nearest neighbors.} 
 \end{figure}
\begin{table}[b]
\caption{\label{tab:table1}%
Values for exchange and anisotropy constants obtained via neutron scattering experiments \cite{wildes2022magnetic,lanconmonolayerscatter} and calculated in this manuscript via DFT. The values are given in $meV$.
}
\begin{ruledtabular}
\begin{tabular}{cccc}
    Parameter & Ref. \cite{lanconmonolayerscatter} & Ref. \cite{wildes2022magnetic} & DFT\\ \hline
    $ \mathcal{J}_1$ & $1.9$ & $1.3$ & $0.06$\\ 
    $ \mathcal{J}_2$ & $-0.1$ & $-0.1$ & $0.025$\\
    $ \mathcal{J}_3$ & $-6.9$ & $-6.8$& $-6.24$\\ 
    $ \mathcal{J}_{\perp}$ & $0.15$ & $0.15$& - \\
    $ \mathcal{D}_x$ & $0$ & $0$ & 0\\
    $ \mathcal{D}_y$ & $0$ & $ -0.21$ & $-0.034$\\
    $ \mathcal{D}_z$ & $0.3$ & $ 0.01$ & $0.005$\\
\end{tabular}
\end{ruledtabular}
\end{table}
We model the system using a Heisenberg Hamiltonian that incorporates the magnetocrystalline anisotropy, the Zeeman energy from an applied external field, and the exchange interaction up to the next-next nearest neighboring spins. The anisotropy and exchange parameters are determined from first-principles calculations, except for the interlayer exchange.
\subsection{Hamiltonian}
We proceed from the Heisenberg Hamiltonian \cite{lanconmonolayerscatter,wildes2022magnetic}
\begin{equation} \label{1}
    H = -\sum_{j,\sigma} \mathcal{J}_{j,j+\sigma}\Vec{S}_j \cdot \Vec{S}_{j+\sigma}
   -\sum_{n,j} \mathcal{D}_n(S^n_j)^2
  -\sum_j{ \gamma \hbar} \Vec{S_{j}}\cdot\Vec{B}_{0},
\end{equation}
where $\Vec{S}_{j}$ is the spin (in units of $\hbar$) at the lattice point $j$, $\Vec{S}^{(n)}_{j}$ is the component of the spin at lattice point $j$ along the $n$-axis ($n \in {x,y,z}$), $ \mathcal{J}_{j, j + \sigma}$ is the exchange interaction between spins at lattice points $j$ and $j+\sigma$, $\mathcal{D}_{n}$ is the strength of the uniaxial anisotropy constant in the direction normal to the lattice planes along $n$-axis. $\Vec{B}_{0}$ is an external magnetic field that can be applied in any general direction that is in the plane, and $\gamma$ is the gyromagnetic ratio given by $\gamma/2 \pi$ = $28$ GHz/T. Each side of the honeycomb is of length $l_{a}$, and the distance between two atoms on top of each other in different layers is $l_{b}$. In this work, we only look at an external field applied along the $z$-axis and the $x$-axis. The zigzag configuration emerges when taking the exchange interaction up to the next-next-nearest neighbor into account. The resulting Hamiltonian for a magnetic monolayer with exchange interaction up to the next-next-nearest neighbor into account is given by

\begin{equation}
\begin{aligned}
    H_{mono} = & -\sum_{j,\sigma_{1}}  \mathcal{J}_{1}\Vec{S}_j \cdot \Vec{S}_{j+\sigma_{1}}
    -\sum_{j,\sigma_{2}}  \mathcal{J}_{2}\Vec{S}_j \cdot \Vec{S}_{j+\sigma_{2}}  \\
    & -\sum_{j,\sigma_{3}}  \mathcal{J}_{3}\Vec{S}_j \cdot \Vec{S}_{j+\sigma_{3}}
   -\sum_j \mathcal{D}_z(S^z_j)^2 - \sum_j \mathcal{D}_y(S^y_j)^2 \\
   & - \sum_j \mathcal{D}_x(S^x_j)^2 
 -\sum_j \gamma \hbar \Vec{S_{j}}\cdot\Vec{B}_{0},
\end{aligned}
\end{equation}
where $\mathcal{J}_{i}$ is the $i^{th}$ nearest neighbor exchange between spins on lattice sites $j$ and $j+\sigma_{i}$. $\Vec{S}_{j+\sigma_{i}}$ represents the spin of the $i^{th}$ nearest neighbor of $\Vec{S}_{j}$. The numerical values for the exchange coupling and anisotropy found through neutron scattering experiments \cite{wildes2022magnetic, lanconmonolayerscatter} and DFT are given in Table 1. 
\subsection{Landau-Lifshitz equation}
We investigate the spin dynamics by the Landau-Lifshitz (LL) equation \cite{landauOG,stancilprabhakar}
\begin{equation}
    \frac{d \vec{m}_{j}(t)}{dt} = -\gamma \left [ \vec{m}_{j}(t) \times \vec{H}_{eff}(t) \right ],
\end{equation}
where $\vec{m}_{j} = \gamma \hbar \vec{S}_{j} / M_{s}$ \cite{stancilprabhakar, rezende2019introduction} is the local magnetic moment at lattice site $j$. The local magnetic moments are treated as classical vectors with constant magnitude. The effective field $\Vec{H}_{eff}$ for the magnetization in each sublattice is derived by taking the functional derivative \cite{stancilprabhakar, rezende2019introduction}
\begin{equation}
    \vec{H}_{eff, j} = \dfrac{1}{\mu _{0} M_{s}} \dfrac{\partial \epsilon}{\partial |m_{j}|},
\end{equation}
where the energy density per unit cell $\epsilon$ is derived from the Hamiltonian in Eq. (1) by substituting the quantum spin operators with the corresponding classical magnetization amplitudes \cite{stancilprabhakar,rezende2019introduction}
\begin{equation}
\resizebox{0.9\hsize}{!}{$
    \begin{aligned}
    \epsilon/M_{s} = -  \sum_{j, \sigma} J_{j,j+\sigma}\Vec{m}_j \cdot \Vec{m}_{j+\sigma} 
   -  \sum_{j}D_n(m^n_j)^2 - \sum_{j} \Vec{m}_{j}\cdot\Vec{B}_{0}, 
\end{aligned}
$}
\end{equation}
where $M_{s}$ is the saturation magnetization, $J_{i} = \mathcal{J}_{i} S/\gamma \hbar$, $D_{n} = \mathcal{D}_{n} S/\gamma \hbar$, and the index $j$ goes over all lattice sites in the unit cell. \ch{Ni} has two holes in the 3d shell, so its spin $S = M_{s}/\gamma \hbar = 1$.
\par
The LL equation is linear in the spin wave approximation. For a spin aligned along the $z$ axis in its ground state, the dynamic magnetization of its excited state is written in the spin wave approximation as \cite{stancilprabhakar,rezende2019introduction}
\begin{equation}
    \Vec{m}_{j} = m_{j}^{z} \hat{z} + (m_{j}^{x} \hat{x} + m_{j}^{y} \hat{y})e^{i (\omega t - k_{x} x_{j} - k_{y} y_{j} - k_{z} z_{j} )},
\end{equation}
where $\Vec{m}_{j}$ is the magnetization vector, $m^{n}_{j}$ is the component of the magnetization along $n$ axis at lattice site $j$. Furthermore, $k_{n}$ is the wavevector along the $n$ axis. The spatial coordinates $x_{j}$, $y_{j}$, and $z_{j}$ can only take discrete values corresponding to the lattice sites $j$. Using Eq. (6), the LL equation is linearized using standard procedure \cite{rezende2019introduction}.
\par
In the ground state, the spins form two sublattices (see Fig. 1). For a field applied in-plane, the ground state is canted and fully described by the azimuthal angles $\theta_{1}$ and $\theta_{2}$ made with the intermediate-axis by the spins of spin-up sublattices and spin-down sublattices, respectively. The classical energy in the ground state follows from Eq. (5) by assuming that the sublattice magnetizations are constant classical vectors with canting angles $\theta_1$ and $\theta_2$ relative to the z-axis \cite{rezende2019introduction}, is given by
\begin{equation}
\begin{aligned}
    \epsilon_{gs} = & M_{s} [ -D_{x} \cos^2\theta_{1}  -D_x \cos^2\theta_{2} -D_z \sin^2\theta_{1} \\ & - D_z \sin^2\theta_{2} - B_0 \sin\theta_{1}  - B_0 \sin\theta_{2} - J_{f} \\ & -  J_{af}( \sin\theta_{1} \sin\theta_{2} - \cos\theta_{1} \cos\theta_{2}) ].
\end{aligned}
\end{equation}
Here $J_{af} = J_{1} + 4J_{2} + 3J_{3}$, and $J_{f} = 2 J_{1} + 2 J_{2}$. We consider first, the easiest case of the magnetic field applied along the easy-axis. The ground state is found by minimizing the energy in Eq. (7) with respect to the angles $\theta_{1}$ and $\theta_{2}$ \cite{rezende2019introduction}, resulting in the antiferromagnetic state (see Fig. 3 (a)), $\theta_{1} = -\theta_{2} = \pi/2$ for $B_{0} < B_{c} = 2 \sqrt{(D_{z} - D_{x})(D_{x} - D_{z} - J_{af})}$, and the spin-flop state (see Fig. 3 (b)) given by $\theta_{1} = \theta_{2} = \theta_{sf}$, where $\theta_{sf}$ is given by \cite{rezende2019introduction}
\begin{equation}
    \sin\theta_{sf} = \dfrac{B_{0}}{2(D_{x} - D_{z} - J_{af})}.
\end{equation}
When the magnetic field is applied along the intermediate axis, the spins are in a canted state. In the canted state, the spins make an angle $\theta_{c}$ with the easy-axis (see Fig. 3 (c)). The angle $\theta_{c}$ is given by \cite{rezende2019introduction}

\begin{equation}
   \sin\theta_{c} = \dfrac{B_{0}}{2(D_{z} - D_{x} - J_{af})}.
\end{equation}

\begin{figure}[]
 \begin{center}
    \begin{subfigure}{0.15\textwidth}
    \includegraphics[width=\linewidth]{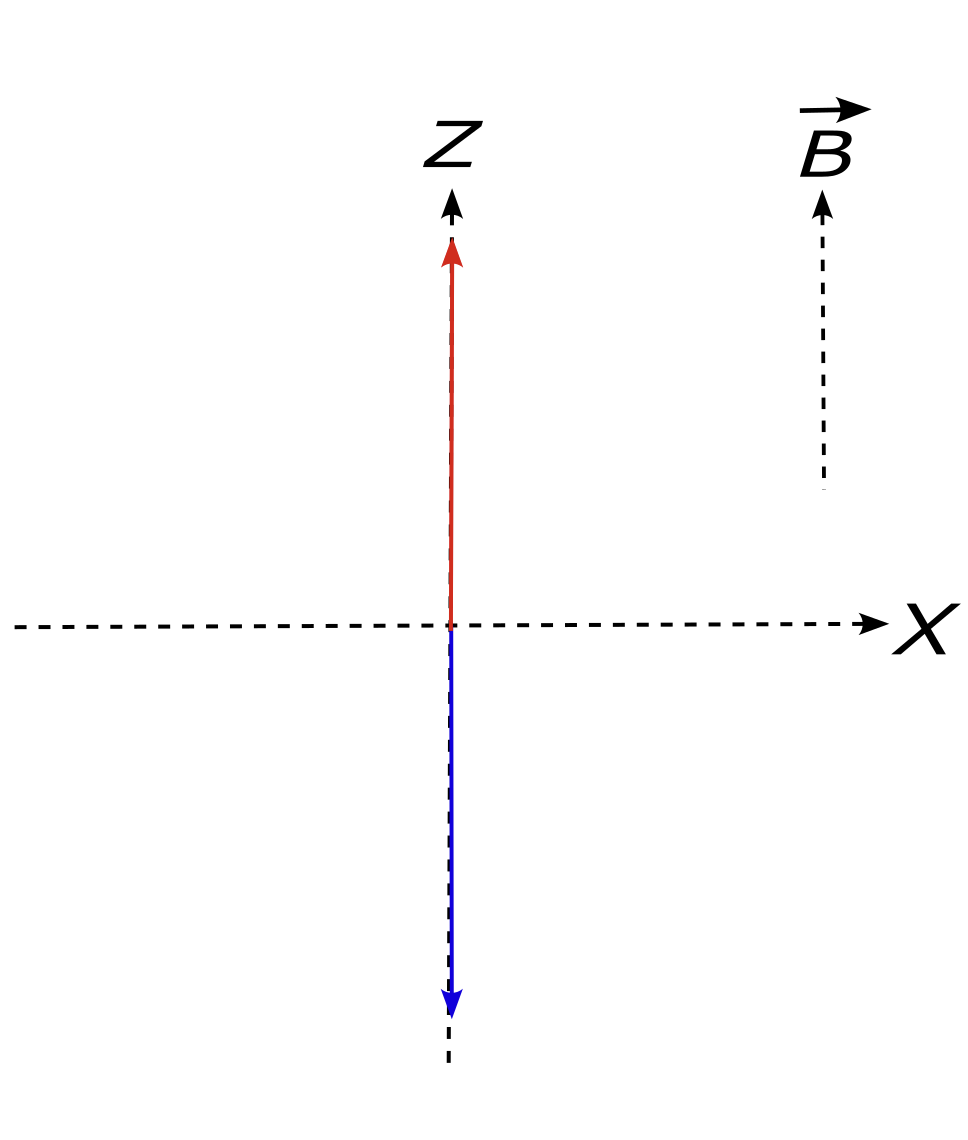}
    \caption{} 
  \end{subfigure}
  \begin{subfigure}{0.143\textwidth}
    \includegraphics[width=\linewidth]{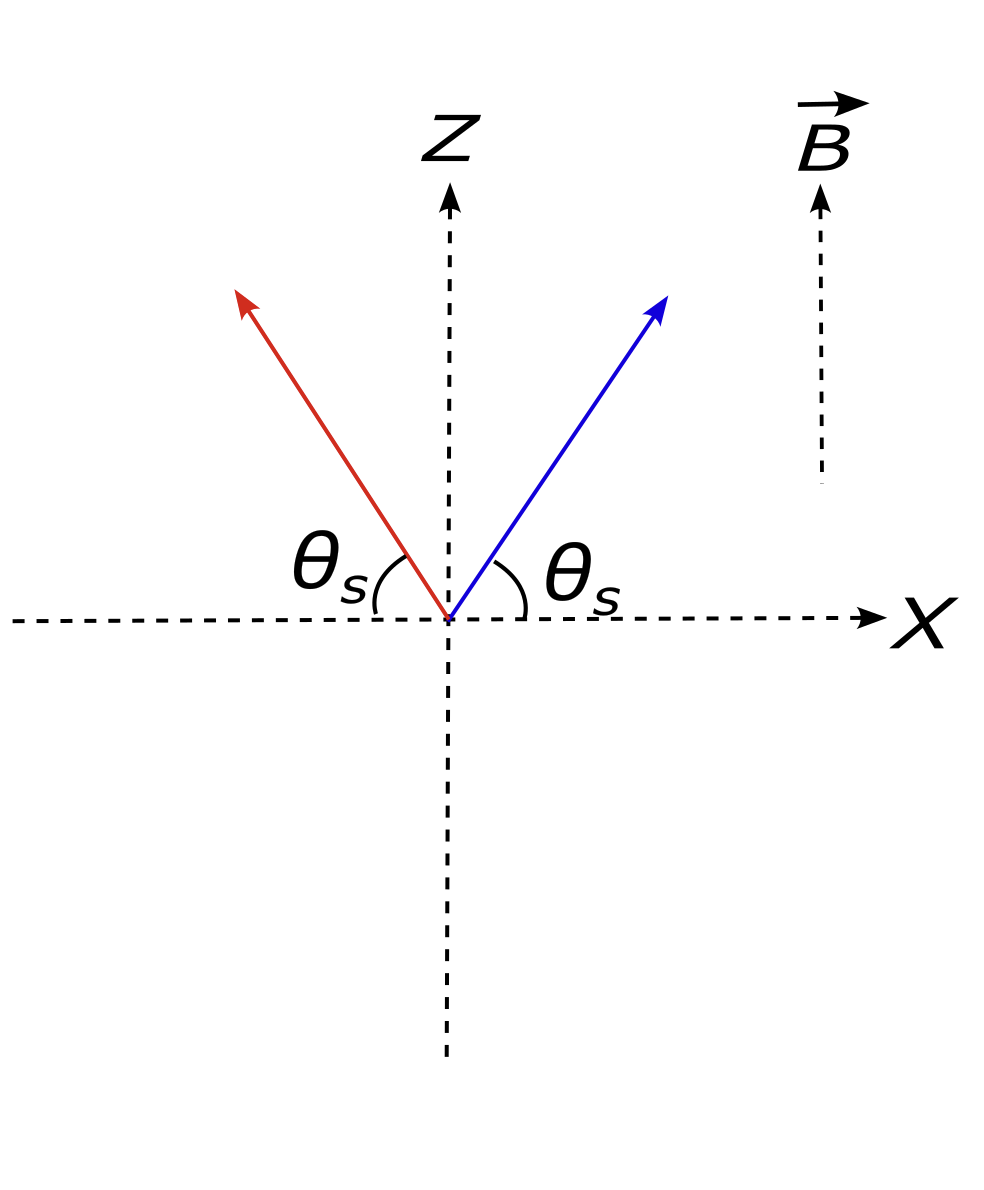}
    \caption{} 
  \end{subfigure}
  \begin{subfigure}{0.155\textwidth}
    \includegraphics[width=\linewidth]{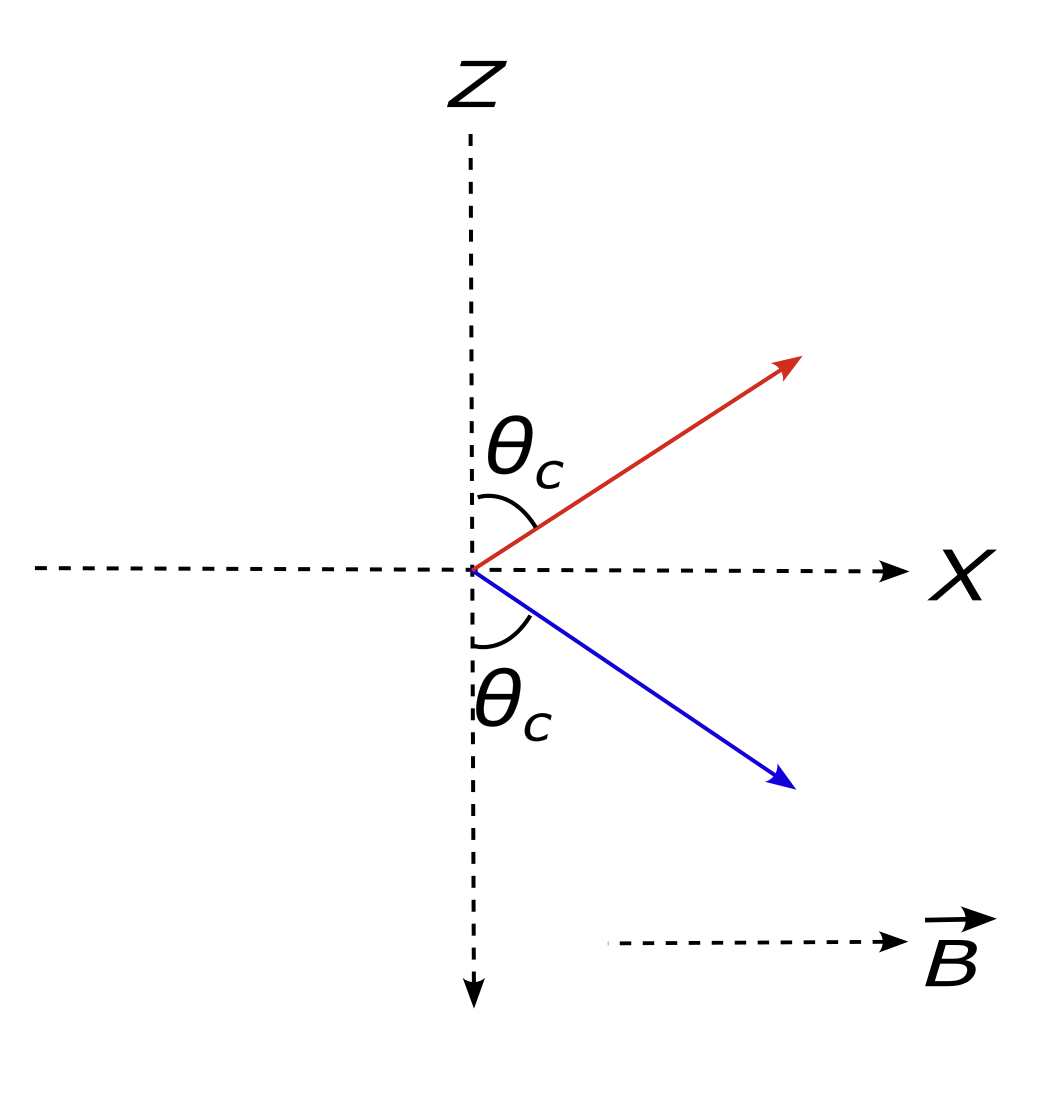}
    \caption{} 
  \end{subfigure}
  \end{center}
    \begin{center}
 \caption{\centerlast Ground state magnetic configurations of an antiferromagnet with 2 sublattices. The easy-axis is along $z$, the $x$-axis is in-plane and perpendicular to $z$-axis. The $y$ axis is out-of-plane. The red arrow indicates the spins in sublattice 1, and the blue arrow indicates the spins in sublattice 2. (a) The antiferromagnetic ground state when no magnetic field, or a weak external magnetic field is applied along the easy-axis. (b) The ground state in the spin flop phase which occurs beyond a critical value for the magnetic field. (c) The magnetic moments aligned in the canted phase which occurs when a magnetic field is applied along the intermediate axis.}
    \end{center}
 \end{figure}

In the next section we describe the first principle methods used to compute the exchange and anisotropy constants for our model.

\section{First Principle Calculations}
\textcolor{black}{We have carried out first-principles simulations of several magnetic phases of monolayer \ch{NiPS3} by the Density Functional Theory code SIESTA \cite{jaime1}. These simulations have been carried out using the PBE GGA exchange-correlation density functional \cite{jaime2}, and include the full spin-orbit interaction self-consistently \cite{jaime3} and optimized pseudopotentials \cite{jaime4} and the triple-zeta double polarized basis set with an energy shift of 10 meV; this a very complete basis having long radia. We used stringent accuracy parameters for the real-space integrals (1000 Rydberg) and reciprocal-space sums (30 x 30 x 1 k points) allowing tolerances for the Hamiltonian and density matrix elements of $10^{-5}$ meV and $10^{-6}$, respectively. We relaxed all inter-atomic forces to below $10^{-3}-10^{-4}$ eV/A depending on the ground state, allowing the unit cell stresses and pressures to below $0.005$ GPa. We have simulated four magnetic phases with spins aligned along the z-axis. These correspond to ferromagnetic and N\'eel antiferromagnetic alignments, as well as the so-called zigzag and strippy phases, where spins are aligned antiferromagnetically along the zigzag/armchair directions and vice versa. No ferromagnetic phase could be converged. The zigzag phase is energetically favoured by about 40 meV over the next most stable state, the N\'eel AFM. \textcolor{black}{In all four phases, the ground state spins ended up completely in-plane.} We have simulated also x- and y-axis-oriented zigzag phases. The x-axis zigzag phase is almost degenerate to the z-axis zigzag phase, while the y-axis zigzag is 0.15 meV higher in energy. These two phases have a slightly higher energy of about 0.5 meV,
reflecting a small spin-orbit-driven magnetic anisotropy. We picked the Hamiltonian of the stable z-axis zigzag phase and used our post-processing tool Grogu \cite{jaime5}, that determines the exchange and intra-atomic anisotropy tensors of a given magnet to any desired neighbor shell using pertubation theory. }
\par
Table I compares the computed exchange and anisotropy constants with experimental values from Refs. \cite{lanconmonolayerscatter} and \cite{wildes2022magnetic}. In agreement with these previous studies, we find a large antiferromagnetic next-next-nearest neighbor exchange, a ferromagnetic and intermediate nearest-neighbor exchange, and a small next-nearest neighbor exchange, resulting in a zigzag spin configuration. As in Ref. \cite{wildes2022magnetic}, we also observe the presence of both an easy-axis and a hard-axis anisotropy. The hard-axis lies out of the plane, which forces the spins to align in-plane, while the weaker easy-axis anisotropy, favors spin alignment along the $z$-direction in the ground state. Our computed values are in closer agreement with the latest experimental results \cite{wildes2022magnetic, nips3bulkneutronscatter}. \textcolor{black}{The experiments also observe that the hard-axis is canted at an angle of $73^{\degree}$, while the DFT calculations find the hard-axis to be completely perpendicular to the $zx$-plane. Since the anisotropy constants are small ($0.01 - 0.1 meV$), it is possible that the interlayer interactions in the bulk \ch{NiPS3} used in the experiments results in the spins lying slightly out-of-plane. This canting of the spins does not result in any qualitative changes and thus can be neglected in the analysis. }The minor differences in the nearest and next-nearest neighbor exchange constants between our calculations and the experimental findings can be attributed to the fact that our simulations focus on monolayers, whereas the experiments were conducted on bulk samples.
\par
In the following section, we present the magnon dispersion computed using the DFT values reported in Table I.
\section{Spin wave dispersion spectra}
Here we solve the linearized LL equation in order to calculate the magnon dispersion in \ch{NiPS3}
for a monolayer, bilayer and in the bulk. We focus on the dispersion for the collinear antiferromagnetic phase.
\subsection{Monolayer}
In the antiferromagnetic phase (Fig. 3 (a)), the spins lie completely in the plane, along the easy-axis. Substituting Eq. (6) in Eq. (4) results in a linear eigenvalue problem\cite{rezende2019introduction,lanconmonolayerscatter}. Solving the eigenvalue problem results in the full magnon dispersion (See Appendix A). We recover the results from Ref. \cite{lanconmonolayerscatter} by disregarding the hard-axis, i.e., $D_{y} = 0$. In this case, the eigenvalues are
\begin{widetext}
\begin{equation}
\resizebox{0.95\hsize}{!}{%
$
\begin{aligned}
\omega_{1 } = B_{0} - \sqrt{A^2-|C| ^2-B^2-|D|^{2}+\sqrt{D^* \left(4 B (A C+B D)+|C|^2 D^*\right)+C^* \left(4 A (A C+B D)-2 C |D|^{2}\right)+D^2 \left(C^*\right)^2}}, \\
\omega_{2 } =  B_{0} + \sqrt{A^2-|C| ^2-B^2-|D|^{2}+\sqrt{D^* \left(4 B (A C+B D)+|C|^2 D^*\right)+C^* \left(4 A (A C+B D)-2 C |D|^{2}\right)+D^2 \left(C^*\right)^2}}, \\
\omega_{3 } =  B_{0} - \sqrt{A^2-|C| ^2-B^2-|D|^{2}-\sqrt{D^* \left(4 B (A C+B D)+|C|^2 D^*\right)+C^* \left(4 A (A C+B D)-2 C |D|^{2}\right)+D^2 \left(C^*\right)^2}}, \\
\omega_{4 } =  B_{0} + \sqrt{A^2-|C| ^2-B^2-|D|^{2}-\sqrt{D^* \left(4 B (A C+B D)+|C|^2 D^*\right)+C^* \left(4 A (A C+B D)-2 C |D|^{2}\right)+D^2 \left(C^*\right)^2}}. 
\end{aligned}
$}
\end{equation}
\end{widetext}
Otherwise, analytical solutions are tedious. However, when $k_{x} = 0$, the eigenvalue problem is easily solvable again and
\begin{widetext}
\begin{equation}
\resizebox{0.95\hsize}{!}{%
$
\begin{split}
\omega_{1 \pm}^{2} = B_{0}^{2}+(A-C) (A_{1}-C)-(B+D)^2 \pm \sqrt{B_{0}^2 (A+A_{1}+2 (B-C+D)) (A+A_{1}-2 (B+C+D))+(A-A_{1})^2 (B+D)^{2}},\\
\omega_{2 \pm}^{2} = B_{0}^{2}+(A+C) (A_{1}+C)-(B-D)^2 \pm \sqrt{B_{0}^2 (A+A_{1}+2 (B+C-D)) (A+A_{1}+2 (C+D-B))+(A-A_{1})^2 (B-D)^{2}}.
\end{split}
$}
\end{equation}
\end{widetext} \hfill \break
\begin{figure}[]
    \centering
    \includegraphics[width=0.48\textwidth]{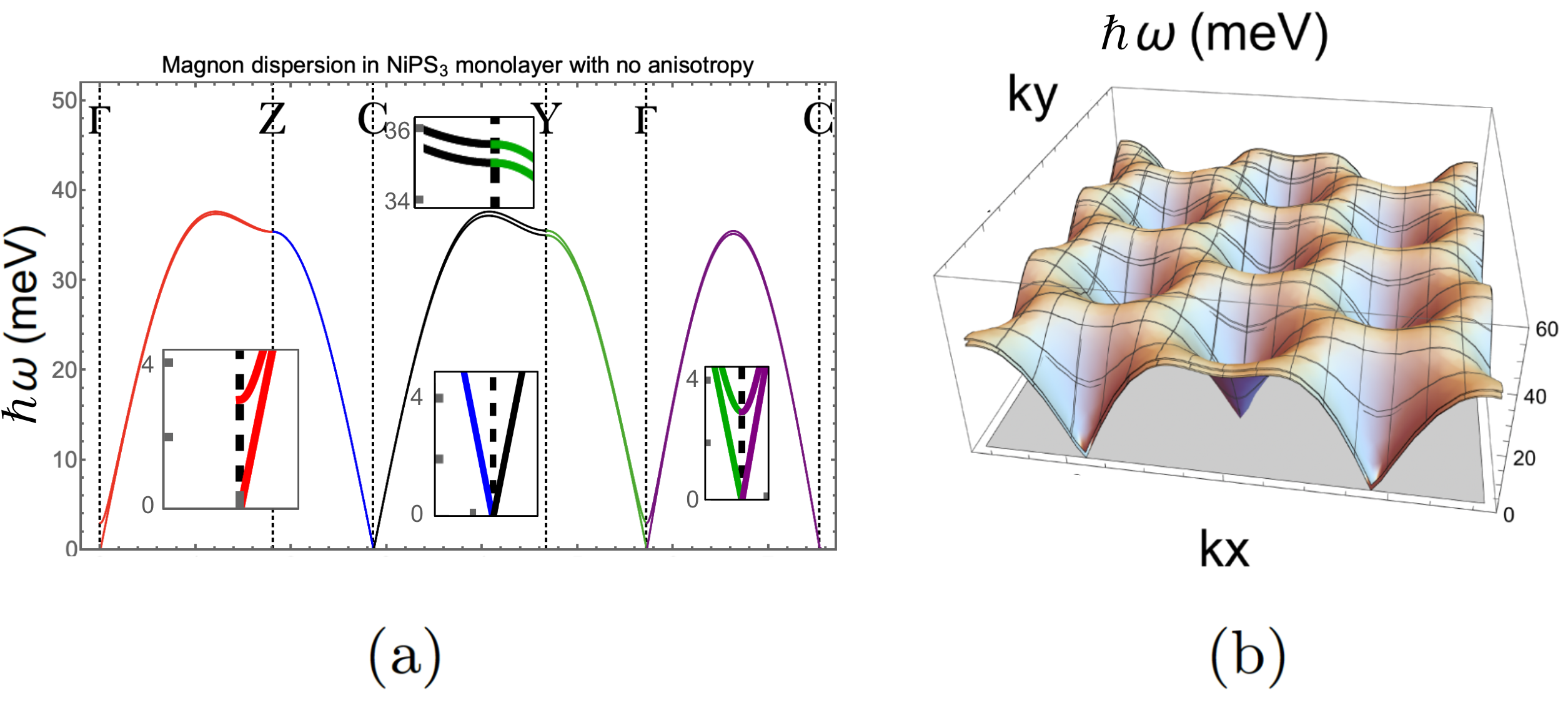}
    \caption{\centerlast Spin wave dispersion in a monolayer \ch{NiPS3} in the absence of anisotropy with only the exchange interactions considered. (a) Line cuts of the dispersion along the relevant paths in the Brillouin zone, also marked in Fig. 1 (b). The insets show blow-ups of the high-symmetry points. (b) 2D dispersion in reciprocal space. The values for exchange interactions are taken from DFT values in Table 1.}
\end{figure}

\begin{figure}[]
    \centering
    \includegraphics[width=0.48\textwidth]{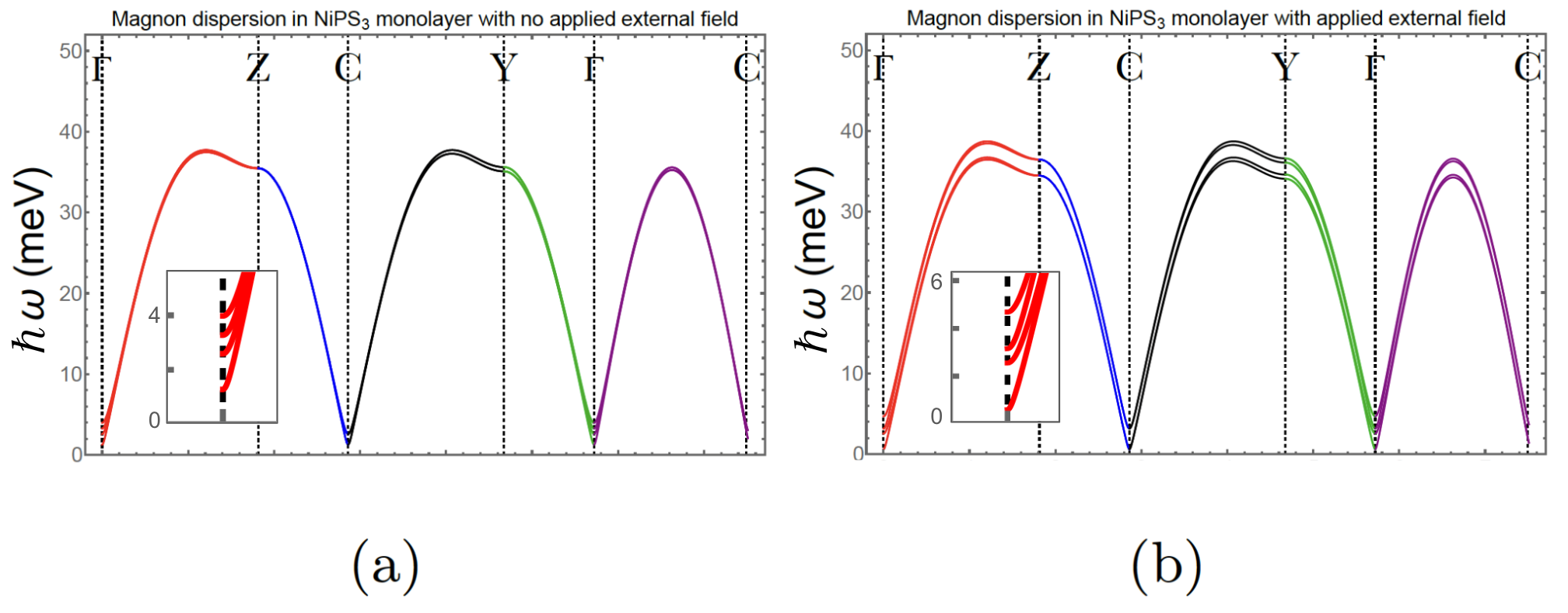}
    \caption{\centerlast (a) Spin wave dispersion in a monolayer \ch{NiPS3} with both easy-axis and hard-axis anisotropy in the absence of an externally applied magnetic field, and (b) in the presence of an externally applied magnetic field of 2T applied along the easy-axis. The hard-axis is taken to be perpendicular to the plane so at equilibrium the spin lies in the plane. The insets show blow-ups of the $\Gamma$ point in each figure. The exchange and anisotropy constants are taken from the DFT values in Table 1.}
\end{figure}

Where
\begin{equation}
\resizebox{0.9\hsize}{!}{$
\begin{aligned}
    A = & 2\left[ D_{z}+J_{1}-2J_{2} \cos \left( \sqrt{3} k_{z} l_{a}\right)-2 J_{2}-3J_{3} \right],  
    A_{1} = & 2(A + D_{y}),\\
    B = & 8 J_{2} \cos \left(\sqrt{3} k_{z}l_{a}/2\right) \cos \left(3 k_{x}l_{a}/2\right),\\
    C = & 4 J_{1} e^{-i k_{x} l_{a}/2} \cos \left(\sqrt{3} k_{z} l_{a}/2\right),\\
    D = & 2\left[e^{i k_{x}l_{a}} \left(J_{1}+2 J_{3} \cos \left( \sqrt{3} k_{z} l_{a}\right)\right)+J_{3} e^{2 i k_{x}l_{a}}\right].
\end{aligned}
$}
\end{equation}

Figure 4 presents the magnon dispersion for monolayer \ch{NiPS3} in the absence of anisotropy or external magnetic fields. Among in-plane antiferromagnetic materials, \ch{NiPS3} is notable for its exceptionally low anisotropy strength \cite{bilsynth}. To establish a baseline, we first analyze the key features of the dispersion in the isotropic limit, before incorporating anisotropy into the system. Figure 4(a) shows two-dimensional cutouts along high-symmetry paths in the Brillouin zone, while the full dispersion across the reciprocal space is depicted in Fig. 4(b). \textcolor{black}{The insets in Fig. 4 (a) show blow-ups of the high-symmetry points. For brevity, we only show blow-ups of the $\Gamma$ points in the other dispersion plots}. The results in Fig. 4 reveal that, in the absence of an external field, the four frequency bands originating from the four sublattices appear as two doubly degenerate bands. \textcolor{black}{The bands form in pairs of higher and lower energy bands. In the higher bands, the up and down spins precess out-of-phase with each other, forming the optical modes. In the lower bands, the up and down spins precess in-phase with each other. The optical and acoustic magnon modes are analogous to the antibonding and bonding orbitals formed in the hydrogen molecule by the destructive and constructive interference of the atomic orbitals respectively.} Additionally, we observe Dirac cones as in other antiferromagnetic materials on a hexagonal lattice. A striking feature of the dispersion is the quadruply degenerate line of Dirac points along the Z-C direction or a Dirac nodal line as predicted for hexagonal zigzag antiferromagnetic systems \cite{topologyzigzag1}. \textcolor{black}{The gap in energy between the top and bottom bands in our calculations is smaller by about $30meV$ at the $\Gamma$ point than that observed in experiments \cite{lanconmonolayerscatter,wildes2022magnetic,nips3bulkneutronscatter}. This discrepancy can be traced to the exchange constants between nearest and next-nearest neighbors, as computed using DFT (see Appendix for the dispersion spectrum using the values from Refs. \cite{wildes2022magnetic,lanconmonolayerscatter}).} 
\par
In Fig. 5, we present the magnon dispersion, incorporating both anisotropy and an external magnetic field. Fig. 5(a) shows the dispersion in the absence of an external field. The degeneracy of the Dirac cones is lifted by the anisotropy. Furthermore, even without an external field, the bands exhibit a small splitting around the points $\Gamma$ and $C$ due to the hard-axis anisotropy, revealing four distinct bands. Along the $Z$-$C$ direction, the quadruply degenerate Dirac nodal line is split into two doubly degenerate Dirac nodal lines.

In Fig. 5(b), we plot the dispersion with anisotropies and a weak external field $(B_{0} < B_{c} \approx 10T)$ applied along the easy-axis. The external field leads to additional splittings or increases the ones induced by the anisotropy. However, along the $Z$-$C$ direction, the two Dirac nodal lines remain doubly degenerate. \textcolor{black}{This persistence suggests that the two doubly degenerate Dirac nodal lines between $Z$ and $C$ are robust against both anisotropy and the external magnetic field. Dirac nodal lines in zigzag antiferromagnets possess topological protection due to a combination of nonsymmorphic and time-reversal symmetries.} \cite{topologyzigzag1}.

The results of prior neutron scattering experiments and numerical calculations in \cite{wildes2022magnetic, nips3bulkneutronscatter} consider the $73.02 \degree$ canting of the hard-axis, which we do not include in our calculations. In spite of this, our results  are qualitatively similar to the results in \cite{wildes2022magnetic, nips3bulkneutronscatter}. In appendix A, we plot the magnon dispersion in monolayer \ch{NiPS3} using the exchange and anisotropy constants from Ref. \cite{wildes2022magnetic}. In this case, our results match closely with the results in Ref. \cite{wildes2022magnetic}. Thus, our model is accurate to the experiments in spite of neglecting the canting of the hard-axis.
\subsection{Bilayer}
The bilayer \ch{NiPS3} \cite{rezende2019introduction, bilsynth, bilayercanting, wildes2022magnetic, nips3bulkneutronscatter} can be thought of as two ferromagnetically coupled antiferromagnetic monolayers with spin Hamiltonian \cite{bilsynth,wildes2022magnetic}
\begin{equation}
    \hat{H}_{bil} = \hat{H^{1}}_{mono} + \hat{H^{2}}_{mono} - \sum_{j, \sigma_{\perp}} \frac{\mathcal{J}_{\perp}}{\hbar^2}\Vec{S}_j \cdot \Vec{S}_{j+\sigma_{\perp}},
\end{equation}
where  $H^{i}_{mono}$ is the Hamiltonian for layer $i \in (1,2)$, $\mathcal{J}_{\perp} > 0$ is the interlayer exchange, $\Vec{S}_j$ is the spin at lattice site $j$, and $\Vec{S}_{j+\sigma_{\perp}}$ denotes the spin at the lattice site directly above or below site $j$. \textcolor{black}{Since DFT calculations of van der Waals interactions are not very accurate, we adopt the value of $J_{\perp}$ from Refs. \cite{lanconmonolayerscatter,wildes2022magnetic}, which is two orders of magnitude smaller than the nearest neighbor exchange constant}. The Landau-Lifshitz (LL) equation for the bilayer becomes a 16$\times$16 matrix equation that we solce numerically. We show the derivation of this matrix in appendix B. Fig. 6 presents the dispersion for bilayer \ch{NiPS3} along high-symmetry paths in the Brillouin zone. The number of independent sublattices and spin bands is 8.

In Fig. 6(a), we show the dispersion for the bilayer in the absence of an external magnetic field. As in the monolayer, the bands split at lower values of $k$ and become nearly degenerate at higher $k$. The bilayer exhibits four doubly degenerate Dirac nodal lines. The interlayer exchange interaction introduces new crossings near $C$, similar to the crossings seen in the monolayer under an applied external field.

In Fig. 6(b), we present the dispersion for bilayer \ch{NiPS3} in the presence of a magnetic field applied along the easy-axis ($B_{0} < B_{c}$) that results in the splitting of spin-degenerate bands. The Dirac nodal lines between $Z$ and $C$ remain doubly degenerate. We also observe additional crossings emerging near $C$ upon the application of the external field.
The dispersion in bulk \ch{NiPS3} \cite{nips3bulkneutronscatter, wildes2022magnetic} can be calculated similarly by replacing $A$ with $A_{b} = A + 2J_{\perp} (1 - \cos{\sqrt{3} k_{y} l_{b})}$ in Eq. (12). We treat the bulk in greater detail in the appendix (See appendix C).
\begin{figure}[]
    \centering
    \includegraphics[width=0.48\textwidth]{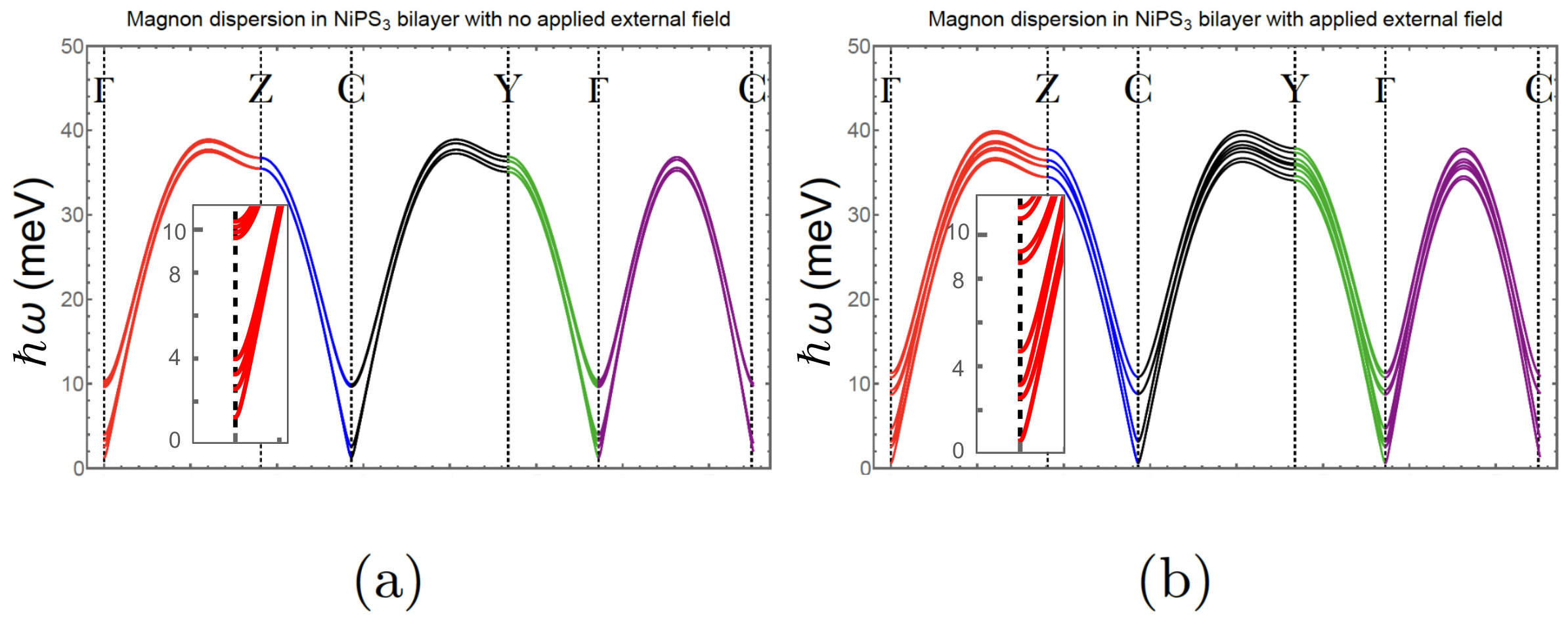}
    \caption{\centerlast Magnon dispersion in a bilayer \ch{NiPS3} in (a) the absence of an externally applied magnetic field, and (b) in the presence of an externally applied magnetic field of 2T (0.25 $meV$) applied along the easy-axis. The intralayer exchange constants and anisotropy constants are taken from DFT values, the interlayer exchange constant is taken from \cite{wildes2022magnetic}. The insets show blow-ups of the $\Gamma$ point in each figure. The values are taken from Table 1.}
\end{figure}
\begin{figure}[]
 \begin{center}
    \begin{subfigure}{0.235\textwidth}
    \includegraphics[width=\linewidth]{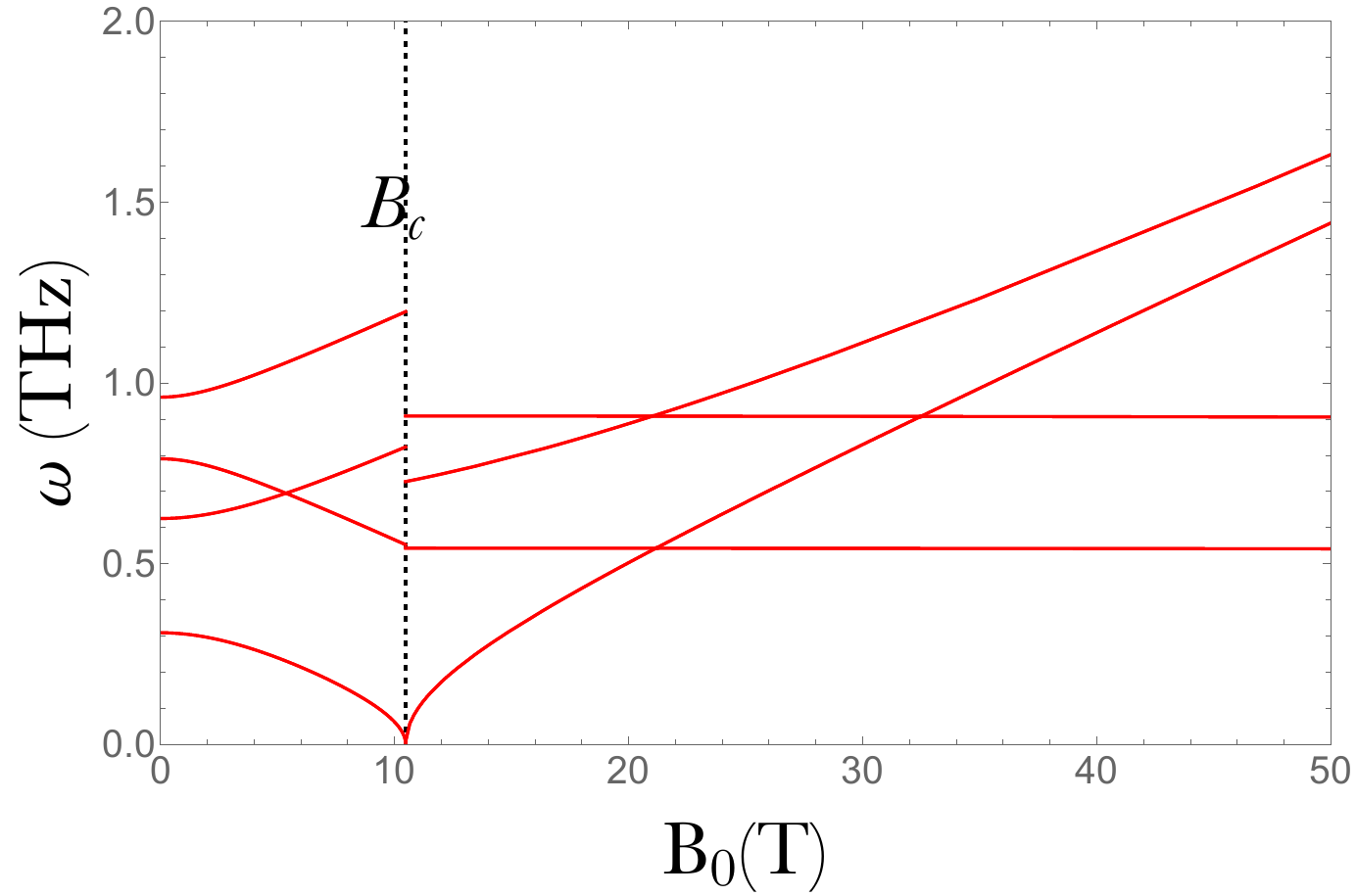}
    \caption{} 
  \end{subfigure}
  \begin{subfigure}{0.235\textwidth}
    \includegraphics[width=\linewidth]{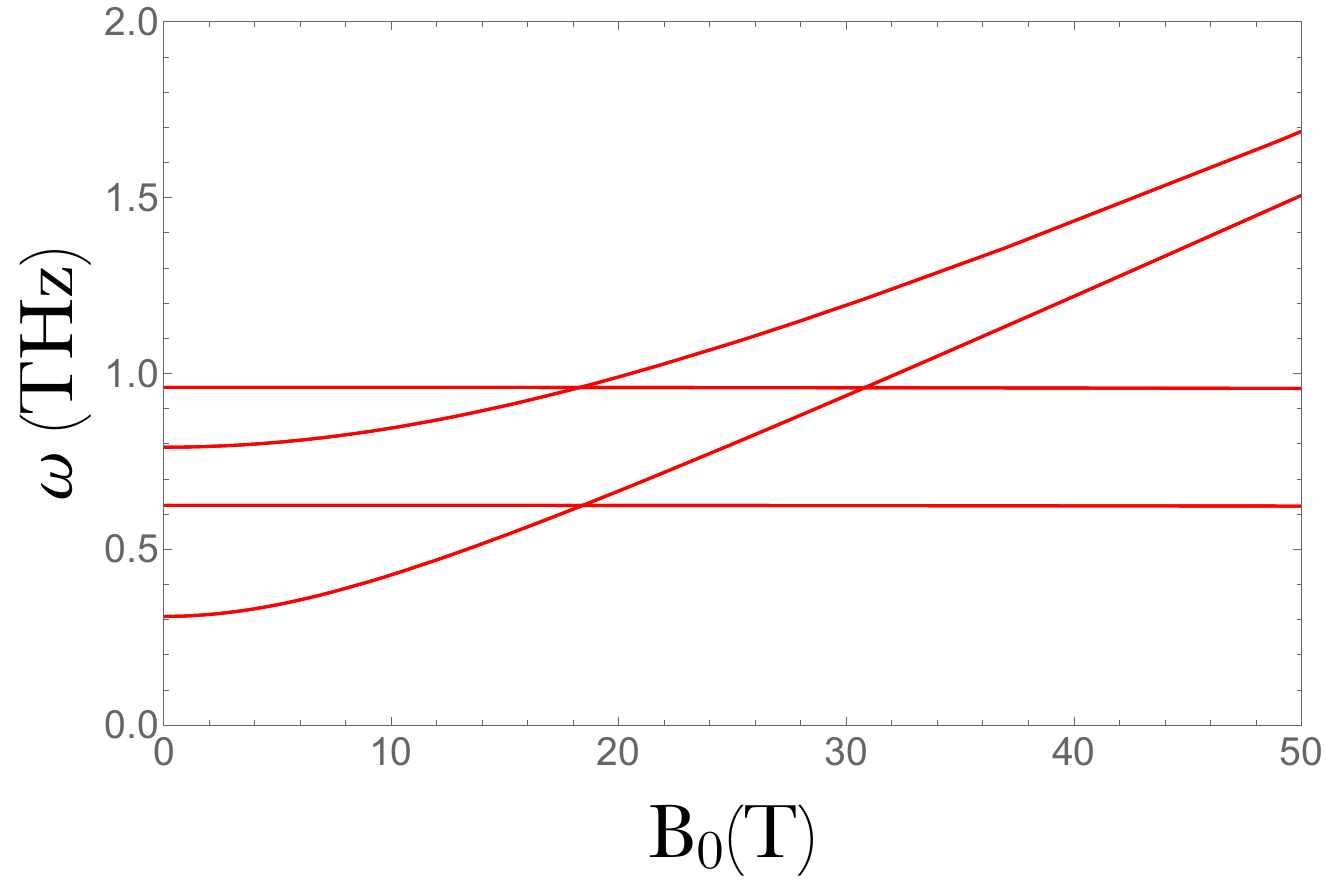}
    \caption{} 
  \end{subfigure}
  
  \end{center}
 \caption{\centerlast AFMR frequency in monolayer \ch{NiPS3} vs. (a) external in-plane magnetic field $B_{0}$ applied along the easy-axis and (b) external in-plane magnetic field $B_{0}$ applied along the intermediate axis. The exchange and anisotropy constants are taken from the DFT values.}
 \end{figure}
In the next section, we compute the evolution of the resonance frequency when the external magnetic field is strong enough to reorient the ground states.
\section{Antiferromagnetic resonance frequency}
Here we compute the antiferromagnetic resonance (AFMR) frequencies \cite{rezende2019introduction,bilsynth}, i.e, the frequencies of the eigenmodes at $k=0$ as a function of an external magnetic field applied in-plane both along, and perpendicular to the easy-axis.
\subsection{Monolayer}
For $B_{0} < B_{c}$ along the N\'eel vector, the spins in the monolayer ground state stay in the collinear antiferromagnetic ground state. When $B_{0} > B_{c}$, the spins transit into the "spin-flop" phase, as described in Sec. II (see Fig. 3(b)). The resonance frequencies in this regime are given by (see appendix D)
\begin{equation}
\resizebox{0.9\hsize}{!}{$
\begin{aligned}
   \omega^{sf}_{1\pm} = & \pm \sqrt{(B_{sf}+A-B_{1}-C-D_{1})(B_{sf}+A_{1}-B-C-D)},\\
   \omega^{sf}_{2\pm} = & \pm \sqrt{(B_{sf}+A+B_{1}+C-D_{1})(B_{sf}+A_{1}+B+C-D)},\\
   \omega^{sf}_{3\pm} = & \pm \sqrt{(B_{sf}+A+B_{1}-C+D_{1})(B_{sf}+A_{1}+B-C+D)},\\
   \omega^{sf}_{4\pm} = & \pm \sqrt{(B_{sf}+A-B_{1}+C+D_{1})(B_{sf}+A_{1}-B+C+D)}.
\end{aligned}
$}
\end{equation}
Where
\begin{equation}
\resizebox{0.8\hsize}{!}{$
\begin{aligned}
    B_{sf} & =  B_{0} \sin \theta_{sf},\\
    A_{sf}  & = 2(D_{z} \sin^{2}\theta_{sf}  + (2 - \cos 2 \theta_{sf})J_{1} \\ & - (4\cos 2 \theta_{sf} - 2)J_{2} \cos \left( \sqrt{3} k_{z} l_{a}\right) \\ & - 3J_{3} \cos 2 \theta_{sf} ),\\
    A_{sf1} & = A_{sf} + 2 D_{y} ,\\
    B = & 2 \left[ J_{2} \cos \left( \sqrt{3} k_{z}l_{a}\right) \cos \left(3 k_{x}l_{a}/2\right)\right],\\ 
    B_{1} = & 4 B \cos 2 \theta_{sf},\\
    C = & 4 J_{1} e^{-i k_{x}l_{a}/2} \cos \left( \sqrt{3} k_{z}l_{a}\right),\\
    D = & 2 \left[e^{i k_{x}l_{a}} \left(J_{1}+2 J_{3} \cos \left( \sqrt{3} k_{z}l_{a}\right)\right)+J_{3} e^{2 i k_{x}l_{a}}\right],\\
    D_{1} = & D \cos 2 \theta_{sf} .
\end{aligned}
$}
\end{equation}
Here $\sin{\theta_{sf}}$ is given in Eq. (8).
\par
In Fig. 7(a), we plot the resonance frequency of monolayer \ch{NiPS3} as a function of the external magnetic field along the easy-axis. The external magnetic field again Zeeman splits the spin degeneracies. At $B_{c}$, the lower band becomes soft, signaling the transition from the antiferromagnetic phase to the spin-flop phase. We calculate $B_{c} \approx$ 10T, which agrees with experimental values \cite{dima}. The higher bands show a discontinuous jump at $B_{c}$ as expected \cite{rezende2019introduction}. Upon increasing the magnetic field to $B_{sat} = 2(D_{x} - D_{z} - J_{af})$, the lowest frequency band once again vanishes and all the spins align along the magnetic field direction (ferromagnetic phase, See Fig. 12 for the AFMR frequency across the full range of $B$). Moreover, we see the bands cross each other at multiple points upon increasing the magnetic field. At these crossings, we expect to see magnon Hanle effect in spin transport \cite{hanle,hanle2}.

The expression for the resonance frequency in the canted state closely resembles that of the spin-flop state. Substituting $\theta_{sf}$ in Eq. (8) with $\theta_{c}$ from Eq. (9), and replacing $\sin^{2}(\theta_{sf}) D_{z}$ with $\cos^{2}(\theta_{c}) D_{z}$ yields the resonance frequency for the canted state. In Fig. 7(b), we show the resonance frequency as a function of an in-plane magnetic field applied perpendicular to the easy-axis. Under this configuration, the spins immediately transition to the canted state with an angle $\theta_{c}$ determined by Eq. (9). As observed previously, the frequency bands increasingly split with increasing magnetic field. At the saturation field, the lowest frequency again goes to zero (see Fig. 12).
\begin{figure}[]
 \begin{center}
    \begin{subfigure}{0.235\textwidth}
    \includegraphics[width=\linewidth]{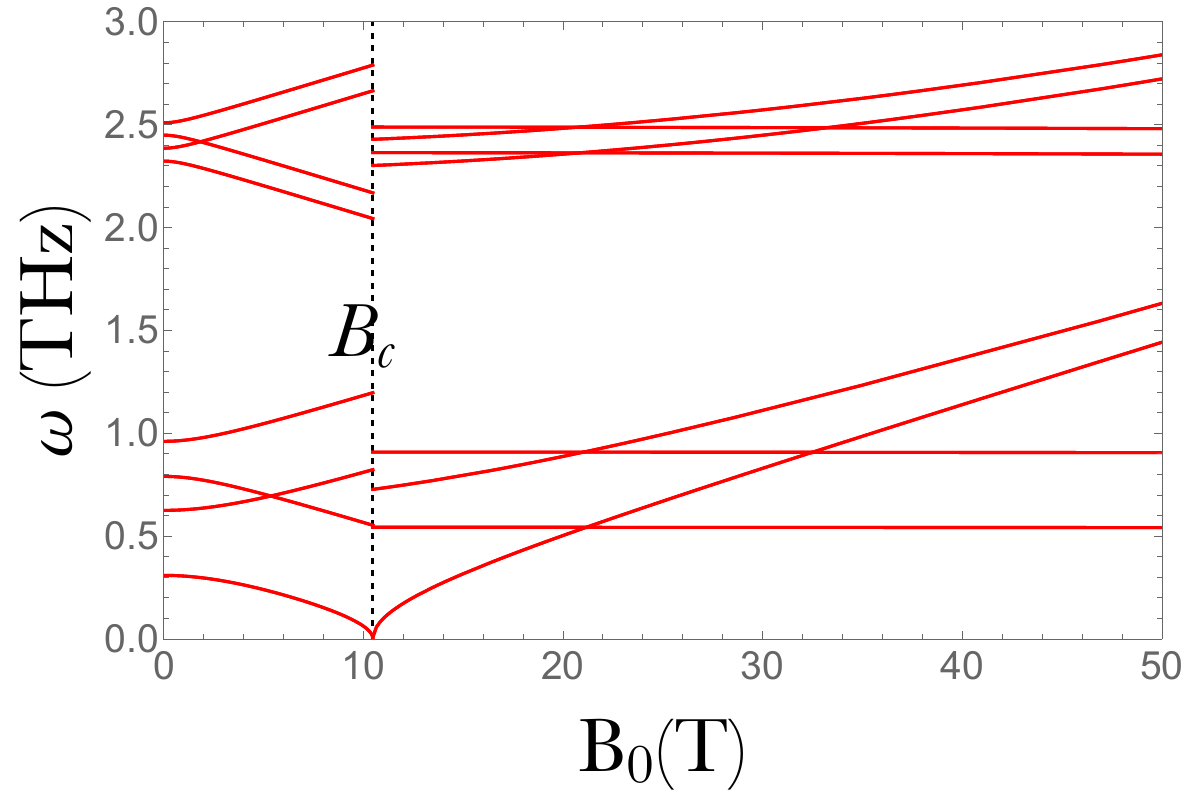}
    \caption{} 
  \end{subfigure}
  \begin{subfigure}{0.235\textwidth}
    \includegraphics[width=\linewidth]{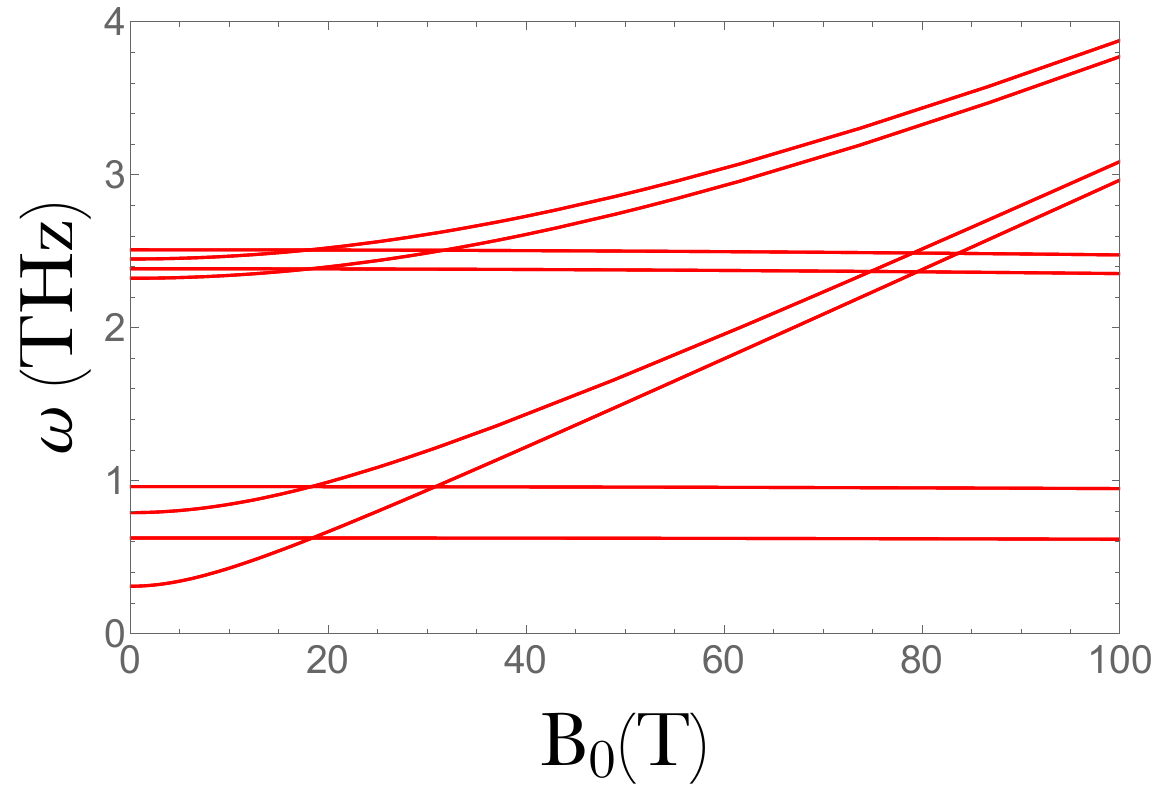}
    \caption{} 
  \end{subfigure}
  \end{center}
 \caption{\centerlast AFMR frequency of bilayer \ch{NiPS3} as a function of (a) external in-plane magnetic field $B_{0}$ along the easy-axis and (b) in-plane magnetic field $B_{0}$ applied along the intermediate axis. The intralayer exchange constants and anisotropy constants are taken from DFT values, the interlayer exchange constant is taken from \cite{wildes2022magnetic}. The values are shown in Table 1.}
 \end{figure}
\subsection{Bilayer}
Finally, we examine the resonance frequency in the bilayer system. In Fig. 8, we plot its resonance frequency as a function of an external magnetic field applied in-plane, along the easy-axis, or perpendicular to the easy-axis. The interlayer exchange interaction between the two layer splits the frequency into eight distinct bands. The overall behavior is similar to that of the monolayer, a few additional crossings are also visible, as shown in Fig. 8.

Fig. 8(a) presents the resonance frequency for bilayer \ch{NiPS3} under an external field applied along the easy-axis. The bands are Zeeman-split by the magnetic field. At $B = B_{c}$, the lowest frequency band reaches zero at the spin-flop transition.

In Fig. 8(b), we show the resonance frequency under an external magnetic field applied perpendicular to the easy-axis. In this case, the spins immediately develop a finite canting angle that increases with the magnetic field. In both the spin-flop and canted states, the splitting introduces multiple new but trivial crossings between the frequency bands. As in the monolayer, when the magnetic field reaches the saturation value, the lowest frequency band reaches zero in both the spin-flop and canted states (see Fig. 12 for the complete AFMR frequency spectrum up to the saturation magnetic field).

\section{Conclusion}
We calculate the magnon dispersion for monolayer and bilayer \ch{NiPS3} based on spin models with \textcolor{black}{exchange and anisotropy constants computed by first-principles that predict a zigzag antiferromagnetic ground state. Our calculations suggest that the next-next-nearest neighbor exchange is the strongest and antiferromagnetic, as is typical of zigzag antiferromagnets. We also identify a strong out-of-plane hard-axis anisotropy and a weaker in-plane easy-axis anisotropy. These findings are in good agreement with previously reported experimental values but do not explain the observed tilting of the hard-axis from the surface normal.} The presence of a hard-axis induces weak band splitting at the $\Gamma$ and $C$ points, a signature of the hard-axis that has been observed in neutron scattering experiments \cite{wildes2022magnetic,nips3bulkneutronscatter}. Additionally, we identify degenerate Dirac nodal lines between the $C$ and $Z$ points, which are robust against anisotropy and external magnetic fields. In the bilayer system, interlayer exchange splits the bands while preserving the degenerate Dirac lines. Furthermore, we investigate the antiferromagnetic resonance (AFMR) frequency as a function of the external magnetic field applied both parallel and perpendicular to the easy-axis, noting significant differences in the behavior of the system across antiferromagnetic, spin-flop, and ferromagnetic states. We find a spin-flop field of around 10T using our DFT values, which agrees with the experimentally observed value of the spin-flop field. We briefly discuss bulk \ch{NiPS3} and relegate the details of the bulk to the appendix.
\par
Our work provides analytical expressions for the AFMR frequency as a function of the magnetic field for both monolayer and bilayer \ch{NiPS3}. Throughout this study, we assume the hard-axis is perpendicular to the crystallographic plane, resulting in spins that are fully confined to the plane in the ground state. This assumption allows for analytical solutions for the dispersion and resonance frequency. Experiments indicate a magnetic structure in which the hard-axis is slightly canted with an angle of 73$\degree$ with the plane. However, the DFT calculations could not reproduce these results. This symmetry breaking would significantly complicate the spin model calculations without much changing the results.
\section{Acknowledgements}
This work reports results of the project "Ronde Open Competitie ENW pakket 21-3" (file number OCENW.M.21.215) which is financed by the Dutch Research Council (NWO). JF has been funded by project PID2022-137078NB-I00 (MCIN/AEI/10.13039/501100011033/FEDER, EU) and the Horizon Europe project TRILMAX, Grant No. 101159646. EVT, was supported by the National Science Center of Poland, project no. UMO-2023/49/B/ST3/02920. GB was supported by JSPS Kakenhi Grants 22H04965 and JP24H02231.

\appendix
\section{EIGENVALUE PROBLEM FOR MONOLAYER}
The unit cell in Fig. 1 contains four independent spins. The eigenvalue problem can be written in the basis
\begin{equation}
    \Vec{m}_{AF} = \begin{pmatrix}
    m_{1a}^{y}&m_{2a}^{y}&m_{1b}^{y}&m_{2b}^{y}&im_{1a}^{z}&im_{2a}^{z}&im_{1b}^{z}&im_{2b}^{z}
    \end{pmatrix}^{T}.
\end{equation}
The sublattices used in the basis are shown in Fig. 1 (b). The eigenvalue matrix relating the frequency to the magnetization vector is given by \cite{lanconmonolayerscatter}
\begin{widetext}
\begin{equation}
    \omega \begin{pmatrix}
        m_{1a}^{x}\\
        m_{2a}^{x}\\
        m_{1b}^{x}\\
        m_{2b}^{x}\\
        im_{1a}^{y}\\
        im_{2a}^{y}\\
        im_{1b}^{y}\\
        im_{2b}^{y}
    \end{pmatrix} = \begin{pmatrix}
        0 & 0 & 0 & 0 & B_{0} + A_{1} & -B & -C & -D\\
        0 & 0 & 0 & 0 & B & B_{0} - A_{1} & D & C\\
        0 & 0 & 0 & 0 & -C^{*} & -D^{*} & B_{0} + A_{1} & -B\\
        0 & 0 & 0 & 0 & D^{*} & C^{*} & B & B_{0} - A_{1}\\
        B_{0} + A & -B & -C & -D & 0 & 0 & 0 & 0\\
        B & B_{0} - A & D & C & 0 & 0 & 0 & 0\\
        -C^{*} & -D^{*} & B_{0} + A & -B & 0 & 0 & 0 & 0\\
        D^{*} & C^{*} & B & B_{0} - A & 0 & 0 & 0 & 0\\
    \end{pmatrix} \begin{pmatrix}
        m_{1a}^{x}\\
        m_{2a}^{x}\\
        m_{1b}^{x}\\
        m_{2b}^{x}\\
        im_{1a}^{y}\\
        im_{2a}^{y}\\
        im_{1b}^{y}\\
        im_{2b}^{y}
    \end{pmatrix}.
\end{equation}
\begin{figure}[]
    \centering
    \includegraphics[width=0.98\textwidth]{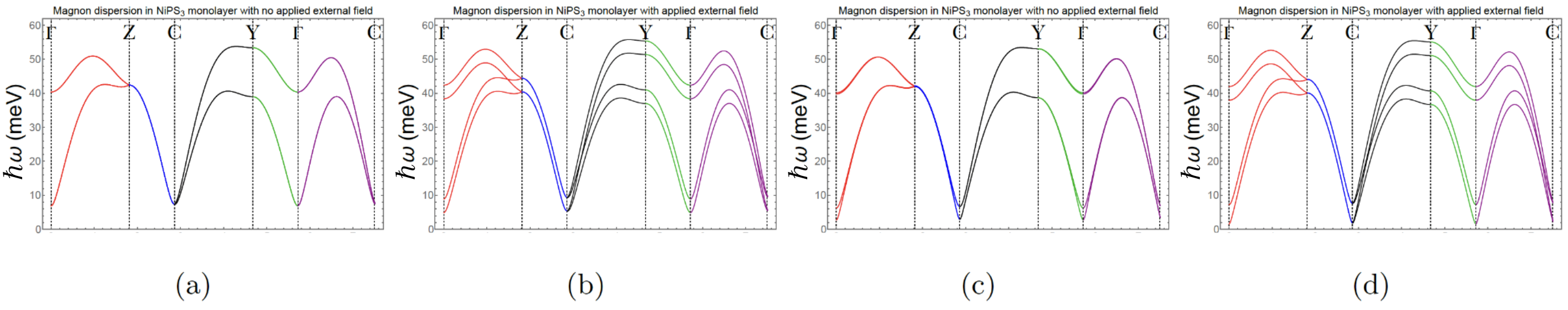}
    \caption{\centerlast Magnon dispersion in a monolayer \ch{NiPS3} with only easy-axis anisotropy \cite{lanconmonolayerscatter} in (a) the absence of an externally applied magnetic field, and (b) in the presence of an externally applied magnetic field of 2T (0.25 $meV$) applied along the easy-axis. (c) Magnon dispersion in a monolayer \ch{NiPS3} with both easy-axis and hard-axis anisotropy \cite{wildes2022magnetic} in the absence of an externally applied magnetic field, and (d) in the presence of an externally applied magnetic field of 2T. The hard-axis is assumed to lie completely out-of-plane such that the constant component of the spin lies completely in-plane.}
\end{figure}
\end{widetext}

 Fig. 9 shows the magnon spectra calculated for the exchange and anisotropy constants given in Ref. \cite{lanconmonolayerscatter, wildes2022magnetic}. Figs. 9 (a-b) show the dispersion for the model with only the easy-axis found in Ref. \cite{lanconmonolayerscatter} and Figs. 9 (c-d) show the dispersion for the model with both an easy, and a hard-axis found in Ref. \cite{wildes2022magnetic}.
\section{EIGENVALUE PROBLEM FOR BILAYER}
The magnetic structure for the primitive unit cell for bilayer \ch{NiPS3} is shown in Fig. 1 (a). The bilayer is described by a unit cell with eight sublattices, four for each layer. For eight sublattices, the resulting eigenvalue problem is given as
\begin{figure}[]
    \centering
    \includegraphics[width=0.45\textwidth]{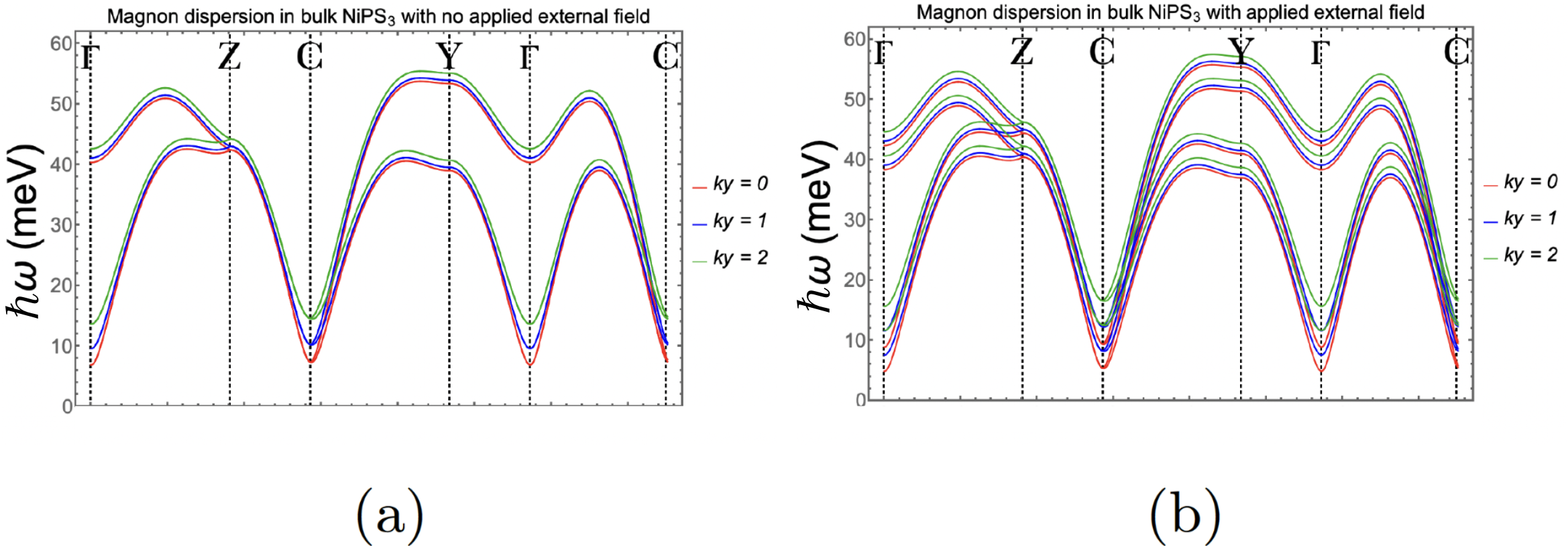}
    \caption{\centerlast Magnon dispersion in bulk \ch{NiPS3} for different values of $k_{y}$ in (a) the absence of an externally applied magnetic field, and (b) in the presence of an externally applied magnetic field of 2T (0.25 $meV$). The exchange constants and anisotropy constants are taken from neutron scattering experiments conducted in \cite{wildes2022magnetic}. The values are taken from Table 1.}
\end{figure}
\begin{widetext}
\begin{equation}
\resizebox{0.95\hsize}{!}{%
$
    \omega \begin{pmatrix}
        m^{x}_{1au}\\
        m^{x}_{2au}\\
        m^{x}_{1bu}\\
        m^{x}_{2bu}\\
        m^{x}_{1ad}\\
        m^{x}_{2ad}\\
        m^{x}_{1bd}\\
        m^{x}_{2bd}\\
        im^{y}_{1au}\\
        im^{y}_{2au}\\
        im^{y}_{1bu}\\
        im^{y}_{2bu}\\
        im^{y}_{1ad}\\
        im^{y}_{2ad}\\
        im^{y}_{1bd}\\
        im^{y}_{2bd}
    \end{pmatrix} = \begin{pmatrix}
        0 & 0 & 0 & 0 & 0 & 0 & 0 & 0 & B_{0} + A_{1b} & -B & -C & -D & -J_{\perp} & 0 & 0 & 0 \\
        0 & 0 & 0 & 0 & 0 & 0 & 0 & 0 & B & B_{0} - A_{1b} & D & C & 0 & J_{\perp} & 0 & 0 \\
        0 & 0 & 0 & 0 & 0 & 0 & 0 & 0 & -C^{*} & -D^{*} & B_{0} + A_{1b} & -B & 0 & 0 & -J_{\perp} & 0\\
        0 & 0 & 0 & 0 & 0 & 0 & 0 & 0 & D^{*} & C^{*} & B & B_{0} - A_{1b} & 0 & 0 & 0 & J_{\perp}\\
        0 & 0 & 0 & 0 & 0 & 0 & 0 & 0 & -J_{\perp} & 0 & 0 & 0 & B_{0} + A_{1b} & -B & -C & -D \\
        0 & 0 & 0 & 0 & 0 & 0 & 0 & 0 & 0 & J_{\perp} & 0 & 0 & B & B_{0} - A_{1b} & D & C \\
    0 & 0 & 0 & 0 &0 &  0 & 0 & 0 & 0 & 0 & -J_{\perp} & 0 & -C^{*} & -D^{*} & B_{0} + A_{1b} & -B \\
        0 & 0 & 0 & 0 &0 & 0 & 0 & 0 & 0 & 0 & 0 & J_{\perp} &  D^{*} & C^{*} & B & B_{0} - A_{1b} \\
        B_{0} + A_{b} & -B & -C & -D & -J_{\perp} & 0 & 0 & 0 & 0 & 0 & 0 & 0 & 0 & 0 & 0 & 0\\
        B & B_{0} - A_{b} & D & C & 0 & J_{\perp} & 0 & 0 & 0 & 0 & 0 & 0 & 0 & 0 & 0 & 0\\
         -C^{*} & -D^{*} & B_{0} + A_{b} & -B & 0 & 0 & -J_{\perp} & 0 & 0 & 0 & 0 & 0 & 0 & 0 & 0 & 0\\
         D^{*} & C^{*} & B & B_{0} - A_{b} & 0 & 0 & 0 & J_{\perp} & 0 & 0 & 0 & 0 & 0 & 0 & 0 & 0\\
        -J_{\perp} & 0 & 0 & 0 & B_{0} + A_{b} & -B & -C & -D & 0 & 0 & 0 & 0 & 0 & 0 & 0 & 0\\
        0 & J_{\perp} & 0 & 0 & B & B_{0} - A_{b} & D & C & 0 & 0 & 0 & 0 & 0 & 0 & 0 & 0\\
        0 & 0 & J_{\perp} & 0 & -C^{*} & -D^{*} & B_{0} + A_{b} & -B & 0 & 0 & 0 & 0 & 0 & 0 & 0 & 0\\
        0 & 0 & 0 & -J_{\perp} & D^{*} & C^{*} & B & B_{0} - A_{b} & 0 & 0 & 0 & 0 & 0 & 0 & 0 & 0
    \end{pmatrix} \begin{pmatrix}
        m^{x}_{1au}\\
        m^{x}_{2au}\\
        m^{x}_{1bu}\\
        m^{x}_{2bu}\\
        m^{x}_{1ad}\\
        m^{x}_{2ad}\\
        m^{x}_{1bd}\\
        m^{x}_{2bd}\\
        im^{y}_{1au}\\
        im^{y}_{2au}\\
        im^{y}_{1bu}\\
        im^{y}_{2bu}\\
        im^{y}_{1ad}\\
        im^{y}_{2ad}\\
        im^{y}_{1bd}\\
        im^{y}_{2bd}
    \end{pmatrix}
$},
\end{equation}
\end{widetext}
Where $A_{1b} = A_{1} + J_{\perp}$, $A_{b} = A + J_{\perp}$, $J_{\perp} = \dfrac{\mathcal{J}_{k} S}{\gamma \hbar}$.

\section{BULK}
The bulk configuration can be described using the same unit as the monolayer. Thus, the unit cell has four sublattices. The unit cell can be repeated over all directions to span the entire structure.
Fig. 10 shows the magnon dispersion for bulk \ch{NiPS3} over the Brillouin zone in the plane for different $k_{y}$. $k_{y}$ is in units of $1/c$, where $c$ is the lattice constant in the $y$ direction. In Fig. 10 (a), we plot the dispersion for bulk \ch{NiPS3} in the absence of an external magnetic field for different $k_{y}$. In Fig. 10 (b), we plot the dispersion for bulk \ch{NiPS3} in the presence of an external magnetic field for different $k_{y}$. Due to small interlayer exchange, the differences with the bilayer are small.

\section{SPIN-FLOP STATE}
\begin{figure}[]
    \centering
    \includegraphics[width=0.45\textwidth]{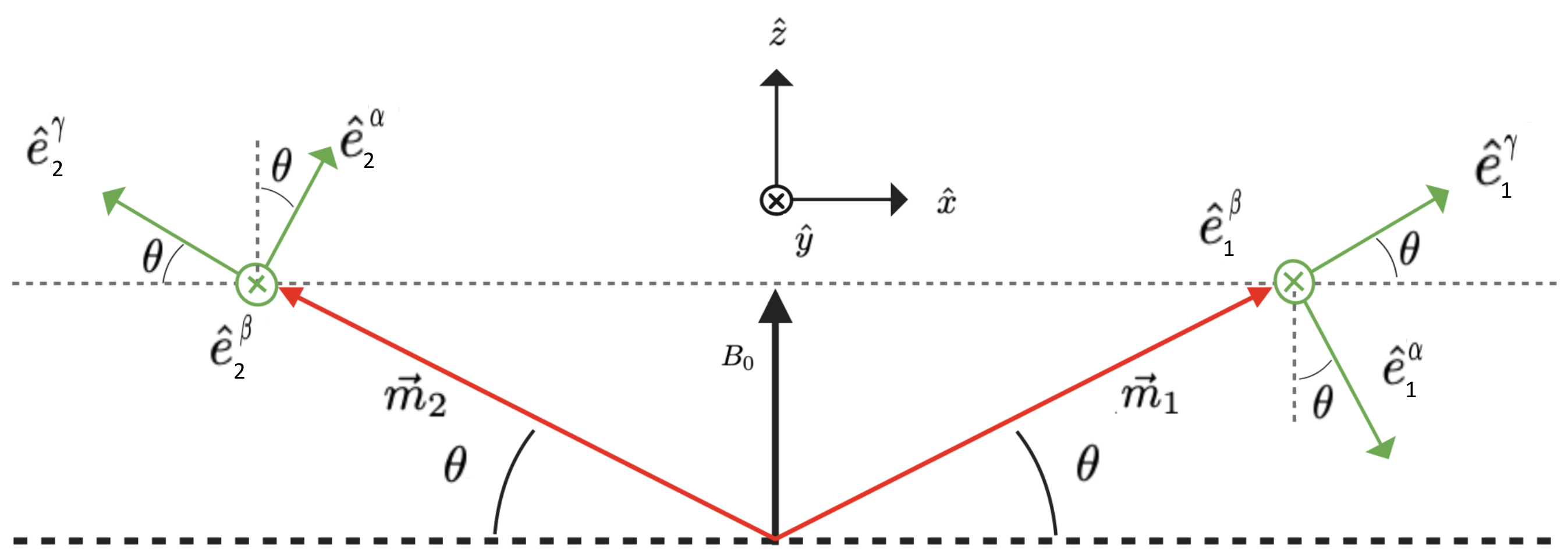}
    \caption{\centerlast Spins in sublattice $\Vec{m}_{1}$ and $\Vec{m}_{2}$. The new coordinate systems are shown with respect to the Cartesian basis. Note that $\Vec{m}_{1 (2)}$ lies exactly along $\hat{e}^{A (B)}_{\gamma}$.}
\end{figure}
\begin{figure}[]
 \begin{center}
    \begin{subfigure}{0.235\textwidth}
    \includegraphics[width=\linewidth]{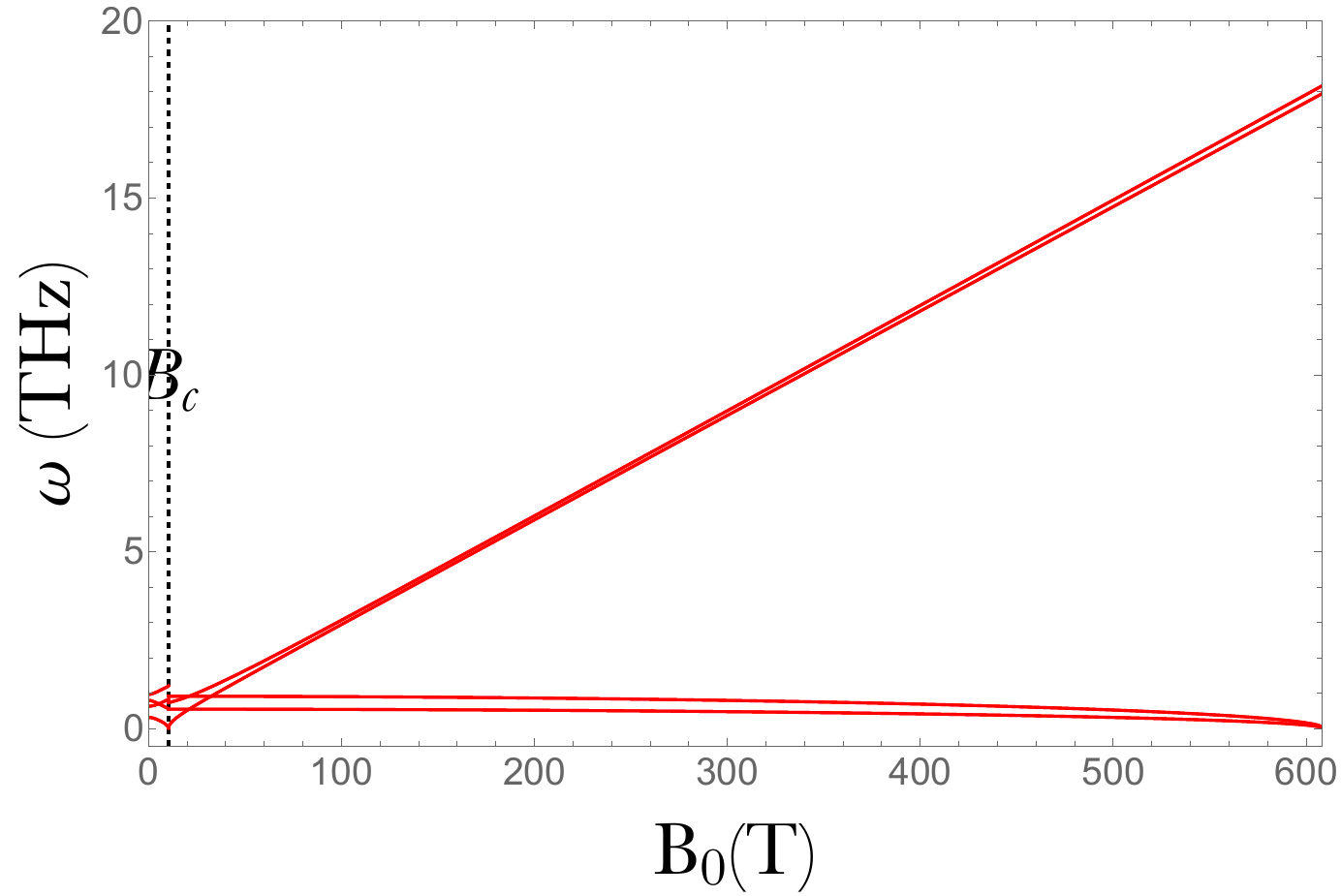}
    \caption{} 
  \end{subfigure}
  \begin{subfigure}{0.235\textwidth}
    \includegraphics[width=\linewidth]{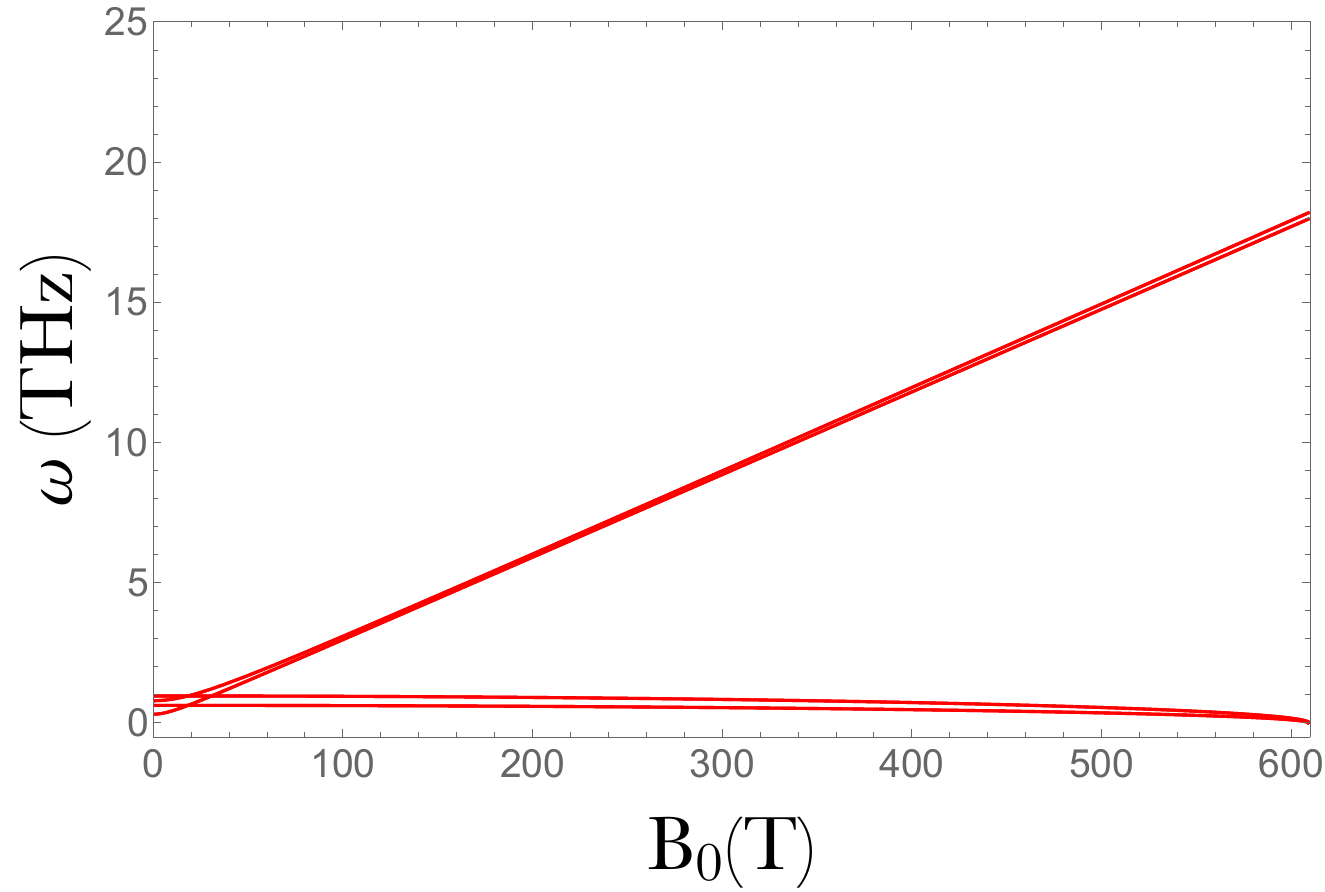}
    \caption{} 
  \end{subfigure}
  \begin{subfigure}{0.235\textwidth}
    \includegraphics[width=\linewidth]{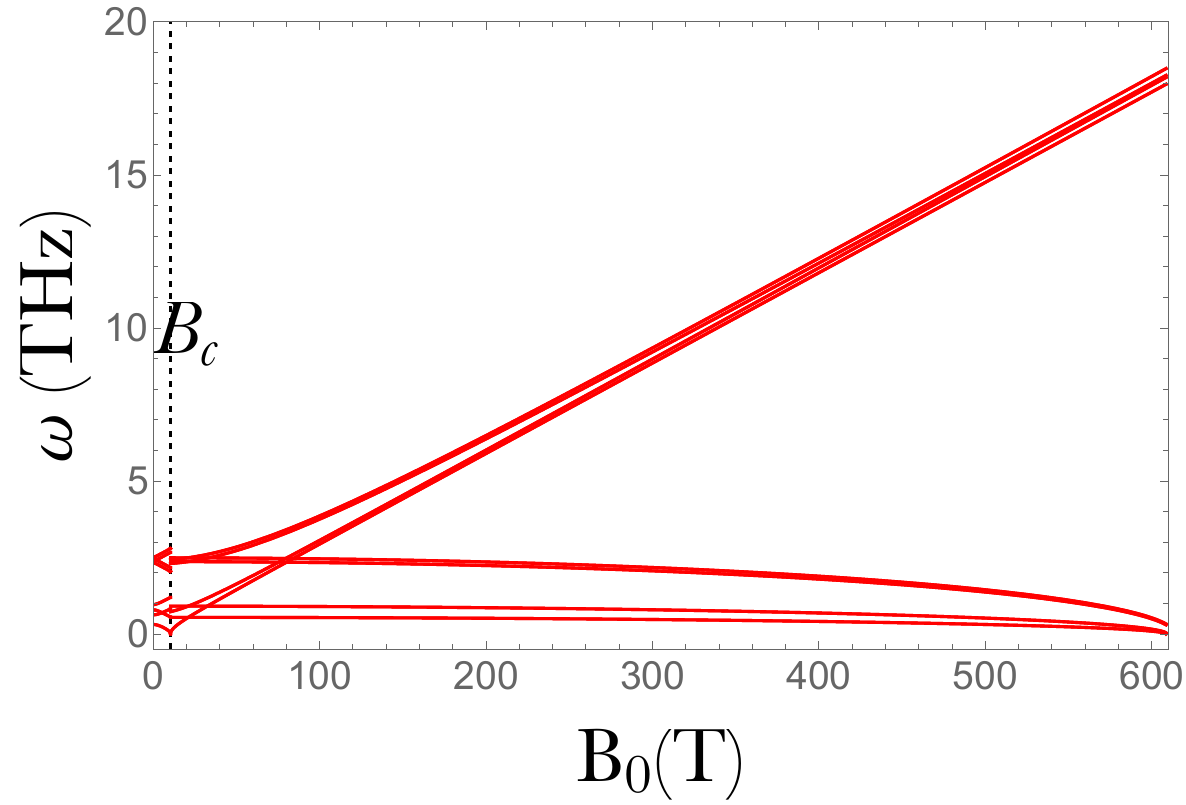}
    \caption{} 
  \end{subfigure}
  \begin{subfigure}{0.235\textwidth}
    \includegraphics[width=\linewidth]{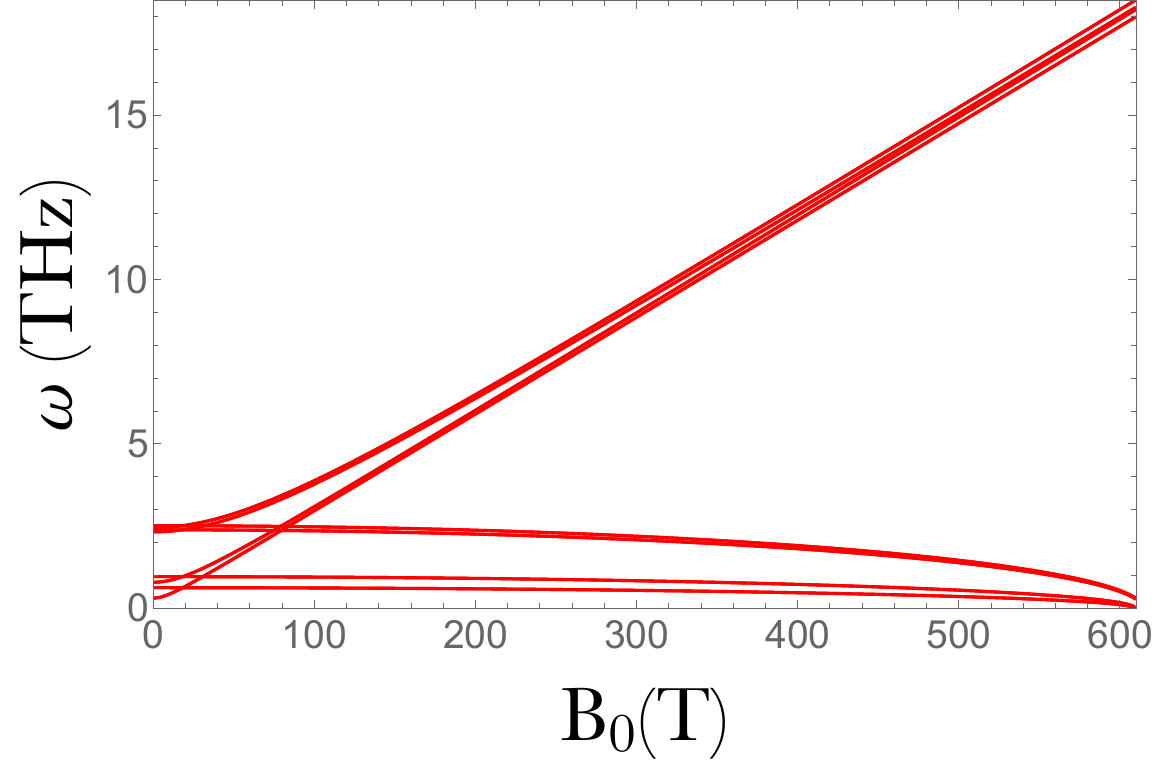}
    \caption{} 
  \end{subfigure}
  \end{center}
 \caption{\centerlast Full AFMR frequency in (a) monolayer \ch{NiPS3} external in-plane magnetic field $B_{0}$ applied along the easy-axis and (b) external in-plane magnetic field $B_{0}$ applied along the intermediate axis. (c) Full AFMR frequency in bilayer \ch{NiPS3} external in-plane magnetic field $B_{0}$ applied along the easy-axis and (d) external in-plane magnetic field $B_{0}$ applied along the intermediate axis. The intralayer exchange constants and anisotropy constants are taken from DFT values, the interlayer exchange constant is taken from \cite{wildes2022magnetic}. The values are shown in Table 1.}
 \end{figure}
In order to describe the excitations of the spin-flop ground state, we define two new coordinate systems such that the component of the spin-up/spin-down magnetization is constant along at least one direction in the new coordinate system. This is a well-known standard procedure for linearizing the LL equation in the case of the spin-flop ground state \cite{rezende2019introduction}. Let $(\hat{e}_{1 (2)}^{\alpha},\hat{e}_{1 (2)}^{\beta},\hat{e}_{1 (2)}^{\gamma})$ be an orthogonal coordinate system such that the component of spins in the spin-up (spin-down) sublattice, denoted by $\Vec{m}_{1 (2)}$ is constant along $\hat{e}_{1 (2)}^{\gamma}$ (See Fig. 11).

The transformations relating the new coordinate system to the cartesian system of coordinates are given by
\begin{equation}
\begin{split}
    \Vec{x} = \sin{\theta} \hat{e}^{\alpha}_{1} + \cos{\theta} \hat{e}^{\gamma}_{1} = \sin{\theta} \hat{e}^{\alpha}_{2} - \cos{\theta} \hat{e}^{\gamma}_{2},\\
    \Vec{y} = \hat{e}^{\beta}_{1} =  \hat{e}^{\gamma}_{1},\\
    \Vec{x} = -\cos{\theta} \hat{e}^{\alpha}_{1} + \sin{\theta} \hat{e}^{\gamma}_{1} = \cos{\theta} \hat{e}^{\alpha}_{2} + \sin{\theta} \hat{e}^{\gamma}_{2}.
\end{split}
\end{equation}
In the new basis, we can write the magnetization as,
\begin{equation}
\begin{split}
    \Vec{m_{1}} = m^{\gamma}_{1} \hat{e}^{\gamma}_{1} + (m^{\alpha}_{1} \hat{e}^{\alpha}_{1} + m^{\beta}_{1} \hat{e}^{\beta}_{1})e^{i (\omega t - k_{x} x - k_{y} y - k_{z} z )},\\
    \Vec{m_{2}} = m^{\gamma}_{2} \hat{e}^{\gamma}_{2} + (m^{\alpha}_{2} \hat{e}^{\alpha}_{2} + m^{\beta}_{2} \hat{e}^{\beta}_{2})e^{i (\omega t - k_{x} x - k_{y} y - k_{z} z )}.\\
\end{split}
\end{equation}
Assuming $m^{\gamma}_{A} = m^{\gamma}_{B} = 1$, we can substitute the Ansatz in Eq. (D2) in Eq. (4) to get the eigenvalue problem relating the frequency to the magnetization in the new basis. We use the sublattices shown in Fig. 1 to form the basis for the eigenvalue problem but in the new coordinate system. The eigenvalue problem in the new basis is
\begin{widetext}
\begin{equation}
\resizebox{0.95\hsize}{!}{%
$
    \omega \begin{pmatrix}
        m_{1a}^{\alpha}\\
        m_{2a}^{\alpha}\\
        m_{1b}^{\alpha}\\
        m_{2b}^{\alpha}\\
        m_{1a}^{\beta}\\
        m_{2a}^{\beta}\\
        m_{1b}^{\beta}\\
        m_{2b}^{\beta}
    \end{pmatrix} = \begin{pmatrix}
        0 & 0 & 0 & 0 &  B_{sf} + A_{sf1} & -B & -C & -D\\
        0 & 0 & 0 & 0 & -B & B_{sf} + A_{sf1} & -D & -C\\
        0 & 0 & 0 & 0 & -C^{*} & -D^{*} & B_{sf} + A_{sf1} & -B\\
        0 & 0 & 0 & 0 & -D^{*} & -C^{*} & -B & B_{sf} + A_{sf1}\\
        B_{sf} + A_{sf} & -B_{1} & -C & -D_{1} & 0 & 0 & 0 & 0\\
        -B_{1} & B_{sf} + A_{sf} & -D_{1} & -C & 0 & 0 & 0 & 0\\
        -C^{*} & -D_{1}^{*} & B_{sf} + A_{sf} & -B_{1} & 0 & 0 & 0 & 0\\
        -D_{1}^{*} & -C^{*} & -B_{1} & B_{sf} + A_{sf} & 0 & 0 & 0 & 0\\
    \end{pmatrix} \begin{pmatrix}
        m_{1a}^{\alpha}\\
        m_{2a}^{\alpha}\\
        m_{1b}^{\alpha}\\
        m_{2b}^{\alpha}\\
        m_{1a}^{\beta}\\
        m_{2a}^{\beta}\\
        m_{1b}^{\beta}\\
        m_{2b}^{\beta}
    \end{pmatrix}.
$}
\end{equation}
\end{widetext}
Diagonalizing Eq. D3 for $k = 0$ gives the full AFMR frequency. Fig. 12 shows the AFMR frequency in a monolayer and bilayer \ch{NiPS3} for the full range of the magnetic field.

\bibliography{apssamp}

\providecommand{\noopsort}[1]{}\providecommand{\singleletter}[1]{#1}%
\begin{thebibliography}{34}%
\makeatletter
\providecommand \@ifxundefined [1]{%
 \@ifx{#1\undefined}
}%
\providecommand \@ifnum [1]{%
 \ifnum #1\expandafter \@firstoftwo
 \else \expandafter \@secondoftwo
 \fi
}%
\providecommand \@ifx [1]{%
 \ifx #1\expandafter \@firstoftwo
 \else \expandafter \@secondoftwo
 \fi
}%
\providecommand \natexlab [1]{#1}%
\providecommand \enquote  [1]{``#1''}%
\providecommand \bibnamefont  [1]{#1}%
\providecommand \bibfnamefont [1]{#1}%
\providecommand \citenamefont [1]{#1}%
\providecommand \href@noop [0]{\@secondoftwo}%
\providecommand \href [0]{\begingroup \@sanitize@url \@href}%
\providecommand \@href[1]{\@@startlink{#1}\@@href}%
\providecommand \@@href[1]{\endgroup#1\@@endlink}%
\providecommand \@sanitize@url [0]{\catcode `\\12\catcode `\$12\catcode `\&12\catcode `\#12\catcode `\^12\catcode `\_12\catcode `\%12\relax}%
\providecommand \@@startlink[1]{}%
\providecommand \@@endlink[0]{}%
\providecommand \url  [0]{\begingroup\@sanitize@url \@url }%
\providecommand \@url [1]{\endgroup\@href {#1}{\urlprefix }}%
\providecommand \urlprefix  [0]{URL }%
\providecommand \Eprint [0]{\href }%
\providecommand \doibase [0]{https://doi.org/}%
\providecommand \selectlanguage [0]{\@gobble}%
\providecommand \bibinfo  [0]{\@secondoftwo}%
\providecommand \bibfield  [0]{\@secondoftwo}%
\providecommand \translation [1]{[#1]}%
\providecommand \BibitemOpen [0]{}%
\providecommand \bibitemStop [0]{}%
\providecommand \bibitemNoStop [0]{.\EOS\space}%
\providecommand \EOS [0]{\spacefactor3000\relax}%
\providecommand \BibitemShut  [1]{\csname bibitem#1\endcsname}%
\let\auto@bib@innerbib\@empty
\bibitem [{\citenamefont {Chumak}\ \emph {et~al.}(2015)\citenamefont {Chumak}, \citenamefont {Vasyuchka}, \citenamefont {Serga},\ and\ \citenamefont {Hillebrands}}]{magnon}%
  \BibitemOpen
  \bibfield  {author} {\bibinfo {author} {\bibfnamefont {A.~V.}\ \bibnamefont {Chumak}}, \bibinfo {author} {\bibfnamefont {V.~I.}\ \bibnamefont {Vasyuchka}}, \bibinfo {author} {\bibfnamefont {A.~A.}\ \bibnamefont {Serga}},\ and\ \bibinfo {author} {\bibfnamefont {B.}~\bibnamefont {Hillebrands}},\ }\bibfield  {title} {\bibinfo {title} {Magnon spintronics},\ }\href@noop {} {\bibfield  {journal} {\bibinfo  {journal} {Nature physics}\ }\textbf {\bibinfo {volume} {11}},\ \bibinfo {pages} {453} (\bibinfo {year} {2015})}\BibitemShut {NoStop}%
\bibitem [{\citenamefont {Rezende}\ and\ \citenamefont {Rezende}(2020)}]{magnonBook}%
  \BibitemOpen
  \bibfield  {author} {\bibinfo {author} {\bibfnamefont {S.~M.}\ \bibnamefont {Rezende}}\ and\ \bibinfo {author} {\bibfnamefont {S.~M.}\ \bibnamefont {Rezende}},\ }\bibfield  {title} {\bibinfo {title} {Magnon spintronics},\ }\href@noop {} {\bibfield  {journal} {\bibinfo  {journal} {Fundamentals of Magnonics}\ ,\ \bibinfo {pages} {287}} (\bibinfo {year} {2020})}\BibitemShut {NoStop}%
\bibitem [{\citenamefont {Prabhakar}\ and\ \citenamefont {Stancil}(2009)}]{stancilprabhakar}%
  \BibitemOpen
  \bibfield  {author} {\bibinfo {author} {\bibfnamefont {A.}~\bibnamefont {Prabhakar}}\ and\ \bibinfo {author} {\bibfnamefont {D.~D.}\ \bibnamefont {Stancil}},\ }\href@noop {} {\emph {\bibinfo {title} {Spin waves: Theory and applications}}},\ Vol.~\bibinfo {volume} {5}\ (\bibinfo  {publisher} {Springer},\ \bibinfo {year} {2009})\BibitemShut {NoStop}%
\bibitem [{\citenamefont {Baghdasaryan}\ and\ \citenamefont {Danoyan}(2017)}]{mewaves}%
  \BibitemOpen
  \bibfield  {author} {\bibinfo {author} {\bibfnamefont {G.}~\bibnamefont {Baghdasaryan}}\ and\ \bibinfo {author} {\bibfnamefont {Z.}~\bibnamefont {Danoyan}},\ }\href@noop {} {\emph {\bibinfo {title} {Magnetoelastic waves}}}\ (\bibinfo  {publisher} {Springer},\ \bibinfo {year} {2017})\BibitemShut {NoStop}%
\bibitem [{\citenamefont {Chumak}\ \emph {et~al.}(2022)\citenamefont {Chumak}, \citenamefont {Kabos}, \citenamefont {Wu}, \citenamefont {Abert}, \citenamefont {Adelmann}, \citenamefont {Adeyeye}, \citenamefont {{\AA}kerman}, \citenamefont {Aliev}, \citenamefont {Anane}, \citenamefont {Awad} \emph {et~al.}}]{comp2}%
  \BibitemOpen
  \bibfield  {author} {\bibinfo {author} {\bibfnamefont {A.~V.}\ \bibnamefont {Chumak}}, \bibinfo {author} {\bibfnamefont {P.}~\bibnamefont {Kabos}}, \bibinfo {author} {\bibfnamefont {M.}~\bibnamefont {Wu}}, \bibinfo {author} {\bibfnamefont {C.}~\bibnamefont {Abert}}, \bibinfo {author} {\bibfnamefont {C.}~\bibnamefont {Adelmann}}, \bibinfo {author} {\bibfnamefont {A.}~\bibnamefont {Adeyeye}}, \bibinfo {author} {\bibfnamefont {J.}~\bibnamefont {{\AA}kerman}}, \bibinfo {author} {\bibfnamefont {F.~G.}\ \bibnamefont {Aliev}}, \bibinfo {author} {\bibfnamefont {A.}~\bibnamefont {Anane}}, \bibinfo {author} {\bibfnamefont {A.}~\bibnamefont {Awad}}, \emph {et~al.},\ }\bibfield  {title} {\bibinfo {title} {Advances in magnetics roadmap on spin-wave computing},\ }\href@noop {} {\bibfield  {journal} {\bibinfo  {journal} {IEEE Transactions on Magnetics}\ }\textbf {\bibinfo {volume} {58}},\ \bibinfo {pages} {1} (\bibinfo {year} {2022})}\BibitemShut {NoStop}%
\bibitem [{\citenamefont {Balynsky}\ \emph {et~al.}(2017)\citenamefont {Balynsky}, \citenamefont {Gutierrez}, \citenamefont {Chiang}, \citenamefont {Kozhevnikov}, \citenamefont {Dudko}, \citenamefont {Filimonov}, \citenamefont {Balandin},\ and\ \citenamefont {Khitun}}]{magnetometer1}%
  \BibitemOpen
  \bibfield  {author} {\bibinfo {author} {\bibfnamefont {M.}~\bibnamefont {Balynsky}}, \bibinfo {author} {\bibfnamefont {D.}~\bibnamefont {Gutierrez}}, \bibinfo {author} {\bibfnamefont {H.}~\bibnamefont {Chiang}}, \bibinfo {author} {\bibfnamefont {A.}~\bibnamefont {Kozhevnikov}}, \bibinfo {author} {\bibfnamefont {G.}~\bibnamefont {Dudko}}, \bibinfo {author} {\bibfnamefont {Y.}~\bibnamefont {Filimonov}}, \bibinfo {author} {\bibfnamefont {A.}~\bibnamefont {Balandin}},\ and\ \bibinfo {author} {\bibfnamefont {A.}~\bibnamefont {Khitun}},\ }\bibfield  {title} {\bibinfo {title} {A magnetometer based on a spin wave interferometer},\ }\href@noop {} {\bibfield  {journal} {\bibinfo  {journal} {Scientific Reports}\ }\textbf {\bibinfo {volume} {7}},\ \bibinfo {pages} {11539} (\bibinfo {year} {2017})}\BibitemShut {NoStop}%
\bibitem [{\citenamefont {Balinskiy}\ \emph {et~al.}(2020)\citenamefont {Balinskiy}, \citenamefont {Chiang}, \citenamefont {Kozhevnikov}, \citenamefont {Filimonov}, \citenamefont {Balandin},\ and\ \citenamefont {Khitun}}]{magnetometer2}%
  \BibitemOpen
  \bibfield  {author} {\bibinfo {author} {\bibfnamefont {M.}~\bibnamefont {Balinskiy}}, \bibinfo {author} {\bibfnamefont {H.}~\bibnamefont {Chiang}}, \bibinfo {author} {\bibfnamefont {A.}~\bibnamefont {Kozhevnikov}}, \bibinfo {author} {\bibfnamefont {Y.}~\bibnamefont {Filimonov}}, \bibinfo {author} {\bibfnamefont {A.}~\bibnamefont {Balandin}},\ and\ \bibinfo {author} {\bibfnamefont {A.}~\bibnamefont {Khitun}},\ }\bibfield  {title} {\bibinfo {title} {A spin-wave magnetometer with a positive feedback},\ }\href@noop {} {\bibfield  {journal} {\bibinfo  {journal} {Journal of Magnetism and Magnetic Materials}\ }\textbf {\bibinfo {volume} {514}},\ \bibinfo {pages} {167046} (\bibinfo {year} {2020})}\BibitemShut {NoStop}%
\bibitem [{\citenamefont {Bauer}\ \emph {et~al.}(2012)\citenamefont {Bauer}, \citenamefont {Saitoh},\ and\ \citenamefont {Van~Wees}}]{calori1}%
  \BibitemOpen
  \bibfield  {author} {\bibinfo {author} {\bibfnamefont {G.~E.}\ \bibnamefont {Bauer}}, \bibinfo {author} {\bibfnamefont {E.}~\bibnamefont {Saitoh}},\ and\ \bibinfo {author} {\bibfnamefont {B.~J.}\ \bibnamefont {Van~Wees}},\ }\bibfield  {title} {\bibinfo {title} {Spin caloritronics},\ }\href@noop {} {\bibfield  {journal} {\bibinfo  {journal} {Nature materials}\ }\textbf {\bibinfo {volume} {11}},\ \bibinfo {pages} {391} (\bibinfo {year} {2012})}\BibitemShut {NoStop}%
\bibitem [{\citenamefont {Yu}\ \emph {et~al.}(2017)\citenamefont {Yu}, \citenamefont {Brechet},\ and\ \citenamefont {Ansermet}}]{calori2}%
  \BibitemOpen
  \bibfield  {author} {\bibinfo {author} {\bibfnamefont {H.}~\bibnamefont {Yu}}, \bibinfo {author} {\bibfnamefont {S.~D.}\ \bibnamefont {Brechet}},\ and\ \bibinfo {author} {\bibfnamefont {J.}~\bibnamefont {Ansermet}},\ }\bibfield  {title} {\bibinfo {title} {Spin caloritronics, origin and outlook},\ }\href@noop {} {\bibfield  {journal} {\bibinfo  {journal} {Physics Letters A}\ }\textbf {\bibinfo {volume} {381}},\ \bibinfo {pages} {825} (\bibinfo {year} {2017})}\BibitemShut {NoStop}%
\bibitem [{\citenamefont {Uchida}(2021)}]{calori3}%
  \BibitemOpen
  \bibfield  {author} {\bibinfo {author} {\bibfnamefont {K.}~\bibnamefont {Uchida}},\ }\bibfield  {title} {\bibinfo {title} {Transport phenomena in spin caloritronics},\ }\href@noop {} {\bibfield  {journal} {\bibinfo  {journal} {Proceedings of the Japan Academy, Series B}\ }\textbf {\bibinfo {volume} {97}},\ \bibinfo {pages} {69} (\bibinfo {year} {2021})}\BibitemShut {NoStop}%
\bibitem [{\citenamefont {Mahmoud}\ \emph {et~al.}(2020)\citenamefont {Mahmoud}, \citenamefont {Ciubotaru}, \citenamefont {Vanderveken}, \citenamefont {Chumak}, \citenamefont {Hamdioui}, \citenamefont {Adelmann},\ and\ \citenamefont {Cotofana}}]{comp1}%
  \BibitemOpen
  \bibfield  {author} {\bibinfo {author} {\bibfnamefont {A.}~\bibnamefont {Mahmoud}}, \bibinfo {author} {\bibfnamefont {F.}~\bibnamefont {Ciubotaru}}, \bibinfo {author} {\bibfnamefont {F.}~\bibnamefont {Vanderveken}}, \bibinfo {author} {\bibfnamefont {A.~V.}\ \bibnamefont {Chumak}}, \bibinfo {author} {\bibfnamefont {S.}~\bibnamefont {Hamdioui}}, \bibinfo {author} {\bibfnamefont {C.}~\bibnamefont {Adelmann}},\ and\ \bibinfo {author} {\bibfnamefont {S.}~\bibnamefont {Cotofana}},\ }\bibfield  {title} {\bibinfo {title} {Introduction to spin wave computing},\ }\href@noop {} {\bibfield  {journal} {\bibinfo  {journal} {Journal of Applied Physics}\ }\textbf {\bibinfo {volume} {128}} (\bibinfo {year} {2020})}\BibitemShut {NoStop}%
\bibitem [{\citenamefont {Burch}\ \emph {et~al.}(2018)\citenamefont {Burch}, \citenamefont {Mandrus},\ and\ \citenamefont {Park}}]{vdw1}%
  \BibitemOpen
  \bibfield  {author} {\bibinfo {author} {\bibfnamefont {K.~S.}\ \bibnamefont {Burch}}, \bibinfo {author} {\bibfnamefont {D.}~\bibnamefont {Mandrus}},\ and\ \bibinfo {author} {\bibfnamefont {J.~G.}\ \bibnamefont {Park}},\ }\bibfield  {title} {\bibinfo {title} {Magnetism in two-dimensional van der waals materials},\ }\href@noop {} {\bibfield  {journal} {\bibinfo  {journal} {Nature}\ }\textbf {\bibinfo {volume} {563}},\ \bibinfo {pages} {47} (\bibinfo {year} {2018})}\BibitemShut {NoStop}%
\bibitem [{\citenamefont {Blei}\ \emph {et~al.}(2021)\citenamefont {Blei}, \citenamefont {Lado}, \citenamefont {Song}, \citenamefont {Dey}, \citenamefont {Erten}, \citenamefont {Pardo}, \citenamefont {Comin}, \citenamefont {Tongay},\ and\ \citenamefont {Botana}}]{vdw2}%
  \BibitemOpen
  \bibfield  {author} {\bibinfo {author} {\bibfnamefont {M.}~\bibnamefont {Blei}}, \bibinfo {author} {\bibfnamefont {J.}~\bibnamefont {Lado}}, \bibinfo {author} {\bibfnamefont {Q.}~\bibnamefont {Song}}, \bibinfo {author} {\bibfnamefont {D.}~\bibnamefont {Dey}}, \bibinfo {author} {\bibfnamefont {O.}~\bibnamefont {Erten}}, \bibinfo {author} {\bibfnamefont {V.}~\bibnamefont {Pardo}}, \bibinfo {author} {\bibfnamefont {R.}~\bibnamefont {Comin}}, \bibinfo {author} {\bibfnamefont {S.}~\bibnamefont {Tongay}},\ and\ \bibinfo {author} {\bibfnamefont {A.}~\bibnamefont {Botana}},\ }\bibfield  {title} {\bibinfo {title} {Synthesis, engineering, and theory of 2{D} van der waals magnets},\ }\href@noop {} {\bibfield  {journal} {\bibinfo  {journal} {Applied Physics Reviews}\ }\textbf {\bibinfo {volume} {8}} (\bibinfo {year} {2021})}\BibitemShut {NoStop}%
\bibitem [{\citenamefont {Yang}\ \emph {et~al.}(2021)\citenamefont {Yang}, \citenamefont {Zhang},\ and\ \citenamefont {Jiang}}]{vdw3}%
  \BibitemOpen
  \bibfield  {author} {\bibinfo {author} {\bibfnamefont {S.}~\bibnamefont {Yang}}, \bibinfo {author} {\bibfnamefont {T.}~\bibnamefont {Zhang}},\ and\ \bibinfo {author} {\bibfnamefont {C.}~\bibnamefont {Jiang}},\ }\bibfield  {title} {\bibinfo {title} {van der waals magnets: Material family, detection and modulation of magnetism, and perspective in spintronics},\ }\href@noop {} {\bibfield  {journal} {\bibinfo  {journal} {Advanced Science}\ }\textbf {\bibinfo {volume} {8}},\ \bibinfo {pages} {2002488} (\bibinfo {year} {2021})}\BibitemShut {NoStop}%
\bibitem [{\citenamefont {Rahman}\ \emph {et~al.}(2021)\citenamefont {Rahman}, \citenamefont {Torres}, \citenamefont {Khan},\ and\ \citenamefont {Lu}}]{vdw4}%
  \BibitemOpen
  \bibfield  {author} {\bibinfo {author} {\bibfnamefont {S.}~\bibnamefont {Rahman}}, \bibinfo {author} {\bibfnamefont {J.~F.}\ \bibnamefont {Torres}}, \bibinfo {author} {\bibfnamefont {A.~R.}\ \bibnamefont {Khan}},\ and\ \bibinfo {author} {\bibfnamefont {Y.}~\bibnamefont {Lu}},\ }\bibfield  {title} {\bibinfo {title} {Recent developments in van der waals antiferromagnetic 2{D} materials: Synthesis, characterization, and device implementation},\ }\href@noop {} {\bibfield  {journal} {\bibinfo  {journal} {ACS {N}ano}\ }\textbf {\bibinfo {volume} {15}},\ \bibinfo {pages} {17175} (\bibinfo {year} {2021})}\BibitemShut {NoStop}%
\bibitem [{\citenamefont {Kim}\ \emph {et~al.}(2019)\citenamefont {Kim}, \citenamefont {Lim}, \citenamefont {Lee}, \citenamefont {Lee}, \citenamefont {Kim}, \citenamefont {Park}, \citenamefont {Jeon}, \citenamefont {Park}, \citenamefont {Park},\ and\ \citenamefont {Cheong}}]{bilsynth}%
  \BibitemOpen
  \bibfield  {author} {\bibinfo {author} {\bibfnamefont {K.}~\bibnamefont {Kim}}, \bibinfo {author} {\bibfnamefont {S.~Y.}\ \bibnamefont {Lim}}, \bibinfo {author} {\bibfnamefont {J.}~\bibnamefont {Lee}}, \bibinfo {author} {\bibfnamefont {S.}~\bibnamefont {Lee}}, \bibinfo {author} {\bibfnamefont {T.~Y.}\ \bibnamefont {Kim}}, \bibinfo {author} {\bibfnamefont {K.}~\bibnamefont {Park}}, \bibinfo {author} {\bibfnamefont {G.~S.}\ \bibnamefont {Jeon}}, \bibinfo {author} {\bibfnamefont {C.-H.}\ \bibnamefont {Park}}, \bibinfo {author} {\bibfnamefont {J.}~\bibnamefont {Park}},\ and\ \bibinfo {author} {\bibfnamefont {H.}~\bibnamefont {Cheong}},\ }\bibfield  {title} {\bibinfo {title} {Suppression of magnetic ordering in {XXZ}-type antiferromagnetic monolayer \ch{NiPS3}},\ }\href@noop {} {\bibfield  {journal} {\bibinfo  {journal} {Nature Communications}\ }\textbf {\bibinfo {volume} {10}},\ \bibinfo {pages} {345} (\bibinfo {year} {2019})}\BibitemShut {NoStop}%
\bibitem [{\citenamefont {Liu}\ \emph {et~al.}(2023)\citenamefont {Liu}, \citenamefont {Zhang}, \citenamefont {Li}, \citenamefont {Li},\ and\ \citenamefont {Pu}}]{modulation}%
  \BibitemOpen
  \bibfield  {author} {\bibinfo {author} {\bibfnamefont {P.}~\bibnamefont {Liu}}, \bibinfo {author} {\bibfnamefont {Y.}~\bibnamefont {Zhang}}, \bibinfo {author} {\bibfnamefont {K.}~\bibnamefont {Li}}, \bibinfo {author} {\bibfnamefont {Y.}~\bibnamefont {Li}},\ and\ \bibinfo {author} {\bibfnamefont {Y.}~\bibnamefont {Pu}},\ }\bibfield  {title} {\bibinfo {title} {Recent advances in 2{D}van der waals magnets: Detection, modulation, and applications},\ }\href@noop {} {\bibfield  {journal} {\bibinfo  {journal} {{S}cience}\ } (\bibinfo {year} {2023})}\BibitemShut {NoStop}%
\bibitem [{\citenamefont {Wildes}\ \emph {et~al.}(2015)\citenamefont {Wildes}, \citenamefont {Simonet}, \citenamefont {Ressouche}, \citenamefont {Mcintyre}, \citenamefont {Avdeev}, \citenamefont {Suard}, \citenamefont {Kimber}, \citenamefont {Lan{\c{c}}on}, \citenamefont {Pepe}, \citenamefont {Moubaraki} \emph {et~al.}}]{wildesstrcture}%
  \BibitemOpen
  \bibfield  {author} {\bibinfo {author} {\bibfnamefont {A.~R.}\ \bibnamefont {Wildes}}, \bibinfo {author} {\bibfnamefont {V.}~\bibnamefont {Simonet}}, \bibinfo {author} {\bibfnamefont {E.}~\bibnamefont {Ressouche}}, \bibinfo {author} {\bibfnamefont {G.~J.}\ \bibnamefont {Mcintyre}}, \bibinfo {author} {\bibfnamefont {M.}~\bibnamefont {Avdeev}}, \bibinfo {author} {\bibfnamefont {E.}~\bibnamefont {Suard}}, \bibinfo {author} {\bibfnamefont {S.~A.}\ \bibnamefont {Kimber}}, \bibinfo {author} {\bibfnamefont {D.}~\bibnamefont {Lan{\c{c}}on}}, \bibinfo {author} {\bibfnamefont {G.}~\bibnamefont {Pepe}}, \bibinfo {author} {\bibfnamefont {B.}~\bibnamefont {Moubaraki}}, \emph {et~al.},\ }\bibfield  {title} {\bibinfo {title} {Magnetic structure of the quasi-two-dimensional antiferromagnet \ch{NiPS3}},\ }\href@noop {} {\bibfield  {journal} {\bibinfo  {journal} {Physical Review B}\ }\textbf {\bibinfo {volume} {92}},\ \bibinfo {pages} {224408} (\bibinfo {year} {2015})}\BibitemShut {NoStop}%
\bibitem [{\citenamefont {Lan{\c{c}}on}\ \emph {et~al.}(2018)\citenamefont {Lan{\c{c}}on}, \citenamefont {Ewings}, \citenamefont {Guidi}, \citenamefont {Formisano},\ and\ \citenamefont {Wildes}}]{lanconmonolayerscatter}%
  \BibitemOpen
  \bibfield  {author} {\bibinfo {author} {\bibfnamefont {D.}~\bibnamefont {Lan{\c{c}}on}}, \bibinfo {author} {\bibfnamefont {R.}~\bibnamefont {Ewings}}, \bibinfo {author} {\bibfnamefont {T.}~\bibnamefont {Guidi}}, \bibinfo {author} {\bibfnamefont {F.}~\bibnamefont {Formisano}},\ and\ \bibinfo {author} {\bibfnamefont {A.}~\bibnamefont {Wildes}},\ }\bibfield  {title} {\bibinfo {title} {Magnetic exchange parameters and anisotropy of the quasi-two-dimensional antiferromagnet \ch{NiPS3}},\ }\href@noop {} {\bibfield  {journal} {\bibinfo  {journal} {Physical Review B}\ }\textbf {\bibinfo {volume} {98}},\ \bibinfo {pages} {134414} (\bibinfo {year} {2018})}\BibitemShut {NoStop}%
\bibitem [{\citenamefont {Wildes}\ \emph {et~al.}(2022)\citenamefont {Wildes}, \citenamefont {Stewart}, \citenamefont {Le}, \citenamefont {Ewings}, \citenamefont {Rule}, \citenamefont {Deng},\ and\ \citenamefont {Anand}}]{wildes2022magnetic}%
  \BibitemOpen
  \bibfield  {author} {\bibinfo {author} {\bibfnamefont {A.}~\bibnamefont {Wildes}}, \bibinfo {author} {\bibfnamefont {J.}~\bibnamefont {Stewart}}, \bibinfo {author} {\bibfnamefont {M.}~\bibnamefont {Le}}, \bibinfo {author} {\bibfnamefont {R.}~\bibnamefont {Ewings}}, \bibinfo {author} {\bibfnamefont {K.}~\bibnamefont {Rule}}, \bibinfo {author} {\bibfnamefont {G.}~\bibnamefont {Deng}},\ and\ \bibinfo {author} {\bibfnamefont {K.}~\bibnamefont {Anand}},\ }\bibfield  {title} {\bibinfo {title} {Magnetic dynamics of \ch{NiPS3}},\ }\href@noop {} {\bibfield  {journal} {\bibinfo  {journal} {Physical Review B}\ }\textbf {\bibinfo {volume} {106}},\ \bibinfo {pages} {174422} (\bibinfo {year} {2022})}\BibitemShut {NoStop}%
\bibitem [{\citenamefont {Scheie}\ \emph {et~al.}(2023)\citenamefont {Scheie}, \citenamefont {Park}, \citenamefont {Villanova}, \citenamefont {Granroth}, \citenamefont {Sarkis}, \citenamefont {Zhang}, \citenamefont {Stone}, \citenamefont {Park}, \citenamefont {Okamoto}, \citenamefont {Berlijn},\ and\ \citenamefont {Tennant}}]{nips3bulkneutronscatter}%
  \BibitemOpen
  \bibfield  {author} {\bibinfo {author} {\bibfnamefont {A.}~\bibnamefont {Scheie}}, \bibinfo {author} {\bibfnamefont {P.}~\bibnamefont {Park}}, \bibinfo {author} {\bibfnamefont {J.~W.}\ \bibnamefont {Villanova}}, \bibinfo {author} {\bibfnamefont {G.~E.}\ \bibnamefont {Granroth}}, \bibinfo {author} {\bibfnamefont {C.~L.}\ \bibnamefont {Sarkis}}, \bibinfo {author} {\bibfnamefont {H.}~\bibnamefont {Zhang}}, \bibinfo {author} {\bibfnamefont {M.~B.}\ \bibnamefont {Stone}}, \bibinfo {author} {\bibfnamefont {J.}~\bibnamefont {Park}}, \bibinfo {author} {\bibfnamefont {S.}~\bibnamefont {Okamoto}}, \bibinfo {author} {\bibfnamefont {T.}~\bibnamefont {Berlijn}},\ and\ \bibinfo {author} {\bibfnamefont {D.~A.}\ \bibnamefont {Tennant}},\ }\bibfield  {title} {\bibinfo {title} {Spin wave hamiltonian and anomalous scattering in \ch{NiPS3}},\ }\href@noop {} {\bibfield  {journal} {\bibinfo  {journal} {Physical Review B}\ }\textbf {\bibinfo {volume} {108}},\ \bibinfo {pages} {104402} (\bibinfo {year} {2023})}\BibitemShut
  {NoStop}%
\bibitem [{\citenamefont {Lee}\ \emph {et~al.}(2018)\citenamefont {Lee}, \citenamefont {Chung}, \citenamefont {Park},\ and\ \citenamefont {Park}}]{topologyzigzag1}%
  \BibitemOpen
  \bibfield  {author} {\bibinfo {author} {\bibfnamefont {K.~H.}\ \bibnamefont {Lee}}, \bibinfo {author} {\bibfnamefont {S.~B.}\ \bibnamefont {Chung}}, \bibinfo {author} {\bibfnamefont {K.}~\bibnamefont {Park}},\ and\ \bibinfo {author} {\bibfnamefont {J.}~\bibnamefont {Park}},\ }\bibfield  {title} {\bibinfo {title} {Magnonic quantum spin hall state in the zigzag and stripe phases of the antiferromagnetic honeycomb lattice},\ }\href@noop {} {\bibfield  {journal} {\bibinfo  {journal} {Physical Review B}\ }\textbf {\bibinfo {volume} {97}},\ \bibinfo {pages} {180401} (\bibinfo {year} {2018})}\BibitemShut {NoStop}%
\bibitem [{\citenamefont {Mermin}\ and\ \citenamefont {Wagner}(1966)}]{merminog}%
  \BibitemOpen
  \bibfield  {author} {\bibinfo {author} {\bibfnamefont {N.~D.}\ \bibnamefont {Mermin}}\ and\ \bibinfo {author} {\bibfnamefont {H.}~\bibnamefont {Wagner}},\ }\bibfield  {title} {\bibinfo {title} {Absence of ferromagnetism or antiferromagnetism in one-or two-dimensional isotropic heisenberg models},\ }\href@noop {} {\bibfield  {journal} {\bibinfo  {journal} {Physical Review Letters}\ }\textbf {\bibinfo {volume} {17}},\ \bibinfo {pages} {1133} (\bibinfo {year} {1966})}\BibitemShut {NoStop}%
\bibitem [{\citenamefont {Mehlawat}\ \emph {et~al.}(2022)\citenamefont {Mehlawat}, \citenamefont {Alfonsov}, \citenamefont {Selter}, \citenamefont {Shemerliuk}, \citenamefont {Aswartham}, \citenamefont {B{\"u}chner},\ and\ \citenamefont {Kataev}}]{bilayercanting}%
  \BibitemOpen
  \bibfield  {author} {\bibinfo {author} {\bibfnamefont {K.}~\bibnamefont {Mehlawat}}, \bibinfo {author} {\bibfnamefont {A.}~\bibnamefont {Alfonsov}}, \bibinfo {author} {\bibfnamefont {S.}~\bibnamefont {Selter}}, \bibinfo {author} {\bibfnamefont {Y.}~\bibnamefont {Shemerliuk}}, \bibinfo {author} {\bibfnamefont {S.}~\bibnamefont {Aswartham}}, \bibinfo {author} {\bibfnamefont {B.}~\bibnamefont {B{\"u}chner}},\ and\ \bibinfo {author} {\bibfnamefont {V.}~\bibnamefont {Kataev}},\ }\bibfield  {title} {\bibinfo {title} {Low-energy excitations and magnetic anisotropy of the layered van der waals antiferromagnet \ch{Ni2P2S6}},\ }\href@noop {} {\bibfield  {journal} {\bibinfo  {journal} {Physical Review B}\ }\textbf {\bibinfo {volume} {105}},\ \bibinfo {pages} {214427} (\bibinfo {year} {2022})}\BibitemShut {NoStop}%
\bibitem [{\citenamefont {Afanasiev}\ \emph {et~al.}(2021)\citenamefont {Afanasiev}, \citenamefont {Hortensius}, \citenamefont {Matthiesen}, \citenamefont {Ma{\~n}as-Valero}, \citenamefont {{\v{S}}i{\v{s}}kins}, \citenamefont {Lee}, \citenamefont {Lesne}, \citenamefont {van Der~Zant}, \citenamefont {Steeneken}, \citenamefont {Ivanov} \emph {et~al.}}]{dima}%
  \BibitemOpen
  \bibfield  {author} {\bibinfo {author} {\bibfnamefont {D.}~\bibnamefont {Afanasiev}}, \bibinfo {author} {\bibfnamefont {J.~R.}\ \bibnamefont {Hortensius}}, \bibinfo {author} {\bibfnamefont {M.}~\bibnamefont {Matthiesen}}, \bibinfo {author} {\bibfnamefont {S.}~\bibnamefont {Ma{\~n}as-Valero}}, \bibinfo {author} {\bibfnamefont {M.}~\bibnamefont {{\v{S}}i{\v{s}}kins}}, \bibinfo {author} {\bibfnamefont {M.}~\bibnamefont {Lee}}, \bibinfo {author} {\bibfnamefont {E.}~\bibnamefont {Lesne}}, \bibinfo {author} {\bibfnamefont {H.~S.}\ \bibnamefont {van Der~Zant}}, \bibinfo {author} {\bibfnamefont {P.~G.}\ \bibnamefont {Steeneken}}, \bibinfo {author} {\bibfnamefont {B.~A.}\ \bibnamefont {Ivanov}}, \emph {et~al.},\ }\bibfield  {title} {\bibinfo {title} {Controlling the anisotropy of a van der waals antiferromagnet with light},\ }\href@noop {} {\bibfield  {journal} {\bibinfo  {journal} {Science Advances}\ }\textbf {\bibinfo {volume} {7}} (\bibinfo {year} {2021})}\BibitemShut {NoStop}%
\bibitem [{\citenamefont {Landau}\ and\ \citenamefont {Lifshitz}(1992)}]{landauOG}%
  \BibitemOpen
  \bibfield  {author} {\bibinfo {author} {\bibfnamefont {L.}~\bibnamefont {Landau}}\ and\ \bibinfo {author} {\bibfnamefont {E.}~\bibnamefont {Lifshitz}},\ }\href@noop {} {\emph {\bibinfo {title} {Perspectives in Theoretical Physics}}}\ (\bibinfo  {publisher} {Elsevier},\ \bibinfo {year} {1992})\ pp.\ \bibinfo {pages} {51--65}\BibitemShut {NoStop}%
\bibitem [{\citenamefont {Rezende}\ \emph {et~al.}(2019)\citenamefont {Rezende}, \citenamefont {Azevedo},\ and\ \citenamefont {Rodr{\'\i}guez-Su{\'a}rez}}]{rezende2019introduction}%
  \BibitemOpen
  \bibfield  {author} {\bibinfo {author} {\bibfnamefont {S.~M.}\ \bibnamefont {Rezende}}, \bibinfo {author} {\bibfnamefont {A.}~\bibnamefont {Azevedo}},\ and\ \bibinfo {author} {\bibfnamefont {R.~L.}\ \bibnamefont {Rodr{\'\i}guez-Su{\'a}rez}},\ }\bibfield  {title} {\bibinfo {title} {Introduction to antiferromagnetic magnons},\ }\href@noop {} {\bibfield  {journal} {\bibinfo  {journal} {Journal of Applied Physics}\ }\textbf {\bibinfo {volume} {126}} (\bibinfo {year} {2019})}\BibitemShut {NoStop}%
\bibitem [{\citenamefont {Soler}\ \emph {et~al.}(2002)\citenamefont {Soler}, \citenamefont {Artacho}, \citenamefont {Gale}, \citenamefont {Garc{\'\i}a}, \citenamefont {Junquera}, \citenamefont {Ordej{\'o}n},\ and\ \citenamefont {S{\'a}nchez-Portal}}]{jaime1}%
  \BibitemOpen
  \bibfield  {author} {\bibinfo {author} {\bibfnamefont {J.~M.}\ \bibnamefont {Soler}}, \bibinfo {author} {\bibfnamefont {E.}~\bibnamefont {Artacho}}, \bibinfo {author} {\bibfnamefont {J.~D.}\ \bibnamefont {Gale}}, \bibinfo {author} {\bibfnamefont {A.}~\bibnamefont {Garc{\'\i}a}}, \bibinfo {author} {\bibfnamefont {J.}~\bibnamefont {Junquera}}, \bibinfo {author} {\bibfnamefont {P.}~\bibnamefont {Ordej{\'o}n}},\ and\ \bibinfo {author} {\bibfnamefont {D.}~\bibnamefont {S{\'a}nchez-Portal}},\ }\bibfield  {title} {\bibinfo {title} {The siesta method for ab initio order-n materials simulation},\ }\href@noop {} {\bibfield  {journal} {\bibinfo  {journal} {Journal of Physics: Condensed Matter}\ }\textbf {\bibinfo {volume} {14}},\ \bibinfo {pages} {2745} (\bibinfo {year} {2002})}\BibitemShut {NoStop}%
\bibitem [{\citenamefont {Perdew}\ \emph {et~al.}(1996)\citenamefont {Perdew}, \citenamefont {Burke},\ and\ \citenamefont {Ernzerhof}}]{jaime2}%
  \BibitemOpen
  \bibfield  {author} {\bibinfo {author} {\bibfnamefont {J.~P.}\ \bibnamefont {Perdew}}, \bibinfo {author} {\bibfnamefont {K.}~\bibnamefont {Burke}},\ and\ \bibinfo {author} {\bibfnamefont {M.}~\bibnamefont {Ernzerhof}},\ }\bibfield  {title} {\bibinfo {title} {Generalized gradient approximation made simple},\ }\href@noop {} {\bibfield  {journal} {\bibinfo  {journal} {Physical Review Letters}\ }\textbf {\bibinfo {volume} {77}},\ \bibinfo {pages} {3865} (\bibinfo {year} {1996})}\BibitemShut {NoStop}%
\bibitem [{\citenamefont {Cuadrado}\ \emph {et~al.}(2021)\citenamefont {Cuadrado}, \citenamefont {Robles}, \citenamefont {Garc{\'\i}a}, \citenamefont {Pruneda}, \citenamefont {Ordej{\'o}n}, \citenamefont {Ferrer},\ and\ \citenamefont {Cerd{\'a}}}]{jaime3}%
  \BibitemOpen
  \bibfield  {author} {\bibinfo {author} {\bibfnamefont {R.}~\bibnamefont {Cuadrado}}, \bibinfo {author} {\bibfnamefont {R.}~\bibnamefont {Robles}}, \bibinfo {author} {\bibfnamefont {A.}~\bibnamefont {Garc{\'\i}a}}, \bibinfo {author} {\bibfnamefont {M.}~\bibnamefont {Pruneda}}, \bibinfo {author} {\bibfnamefont {P.}~\bibnamefont {Ordej{\'o}n}}, \bibinfo {author} {\bibfnamefont {J.}~\bibnamefont {Ferrer}},\ and\ \bibinfo {author} {\bibfnamefont {J.~I.}\ \bibnamefont {Cerd{\'a}}},\ }\bibfield  {title} {\bibinfo {title} {Validity of the on-site spin-orbit coupling approximation},\ }\href@noop {} {\bibfield  {journal} {\bibinfo  {journal} {Physical Review B}\ }\textbf {\bibinfo {volume} {104}},\ \bibinfo {pages} {195104} (\bibinfo {year} {2021})}\BibitemShut {NoStop}%
\bibitem [{\citenamefont {Rivero}\ \emph {et~al.}(2015)\citenamefont {Rivero}, \citenamefont {Garc{\'\i}a-Su{\'a}rez}, \citenamefont {Pere{\~n}iguez}, \citenamefont {Utt}, \citenamefont {Yang}, \citenamefont {Bellaiche}, \citenamefont {Park}, \citenamefont {Ferrer},\ and\ \citenamefont {Barraza-Lopez}}]{jaime4}%
  \BibitemOpen
  \bibfield  {author} {\bibinfo {author} {\bibfnamefont {P.}~\bibnamefont {Rivero}}, \bibinfo {author} {\bibfnamefont {V.~M.}\ \bibnamefont {Garc{\'\i}a-Su{\'a}rez}}, \bibinfo {author} {\bibfnamefont {D.}~\bibnamefont {Pere{\~n}iguez}}, \bibinfo {author} {\bibfnamefont {K.}~\bibnamefont {Utt}}, \bibinfo {author} {\bibfnamefont {Y.}~\bibnamefont {Yang}}, \bibinfo {author} {\bibfnamefont {L.}~\bibnamefont {Bellaiche}}, \bibinfo {author} {\bibfnamefont {K.}~\bibnamefont {Park}}, \bibinfo {author} {\bibfnamefont {J.}~\bibnamefont {Ferrer}},\ and\ \bibinfo {author} {\bibfnamefont {S.}~\bibnamefont {Barraza-Lopez}},\ }\bibfield  {title} {\bibinfo {title} {Systematic pseudopotentials from reference eigenvalue sets for dft calculations},\ }\href@noop {} {\bibfield  {journal} {\bibinfo  {journal} {Computational Materials Science}\ }\textbf {\bibinfo {volume} {98}},\ \bibinfo {pages} {372} (\bibinfo {year} {2015})}\BibitemShut {NoStop}%
\bibitem [{\citenamefont {Mart{\'\i}nez-Carracedo}\ \emph {et~al.}(2023)\citenamefont {Mart{\'\i}nez-Carracedo}, \citenamefont {Oroszl{\'a}ny}, \citenamefont {Garc{\'\i}a-Fuente}, \citenamefont {Ny{\'a}ri}, \citenamefont {Udvardi}, \citenamefont {Szunyogh},\ and\ \citenamefont {Ferrer}}]{jaime5}%
  \BibitemOpen
  \bibfield  {author} {\bibinfo {author} {\bibfnamefont {G.}~\bibnamefont {Mart{\'\i}nez-Carracedo}}, \bibinfo {author} {\bibfnamefont {L.}~\bibnamefont {Oroszl{\'a}ny}}, \bibinfo {author} {\bibfnamefont {A.}~\bibnamefont {Garc{\'\i}a-Fuente}}, \bibinfo {author} {\bibfnamefont {B.}~\bibnamefont {Ny{\'a}ri}}, \bibinfo {author} {\bibfnamefont {L.}~\bibnamefont {Udvardi}}, \bibinfo {author} {\bibfnamefont {L.}~\bibnamefont {Szunyogh}},\ and\ \bibinfo {author} {\bibfnamefont {J.}~\bibnamefont {Ferrer}},\ }\bibfield  {title} {\bibinfo {title} {Relativistic magnetic interactions from nonorthogonal basis sets},\ }\href@noop {} {\bibfield  {journal} {\bibinfo  {journal} {Physical Review B}\ }\textbf {\bibinfo {volume} {108}},\ \bibinfo {pages} {214418} (\bibinfo {year} {2023})}\BibitemShut {NoStop}%
\bibitem [{\citenamefont {G{\"u}ckelhorn}\ \emph {et~al.}(2023)\citenamefont {G{\"u}ckelhorn}, \citenamefont {de-la Pe{\~n}a}, \citenamefont {Scheufele}, \citenamefont {Grammer}, \citenamefont {Opel}, \citenamefont {Gepr{\"a}gs}, \citenamefont {Cuevas}, \citenamefont {Gross}, \citenamefont {Huebl}, \citenamefont {Kamra} \emph {et~al.}}]{hanle}%
  \BibitemOpen
  \bibfield  {author} {\bibinfo {author} {\bibfnamefont {J.}~\bibnamefont {G{\"u}ckelhorn}}, \bibinfo {author} {\bibfnamefont {S.}~\bibnamefont {de-la Pe{\~n}a}}, \bibinfo {author} {\bibfnamefont {M.}~\bibnamefont {Scheufele}}, \bibinfo {author} {\bibfnamefont {M.}~\bibnamefont {Grammer}}, \bibinfo {author} {\bibfnamefont {M.}~\bibnamefont {Opel}}, \bibinfo {author} {\bibfnamefont {S.}~\bibnamefont {Gepr{\"a}gs}}, \bibinfo {author} {\bibfnamefont {J.~C.}\ \bibnamefont {Cuevas}}, \bibinfo {author} {\bibfnamefont {R.}~\bibnamefont {Gross}}, \bibinfo {author} {\bibfnamefont {H.}~\bibnamefont {Huebl}}, \bibinfo {author} {\bibfnamefont {A.}~\bibnamefont {Kamra}}, \emph {et~al.},\ }\href@noop {} {\bibfield  {journal} {\bibinfo  {journal} {Observation of the nonreciprocal magnon Hanle effect, Physical Review Letters}\ }\textbf {\bibinfo {volume} {130}},\ \bibinfo {pages} {216703} (\bibinfo {year} {2023})}\BibitemShut {NoStop}%
\bibitem [{\citenamefont {Wimmer}\ \emph {et~al.}(2020)\citenamefont {Wimmer}, \citenamefont {Kamra}, \citenamefont {G{\"u}ckelhorn}, \citenamefont {Opel}, \citenamefont {Gepr{\"a}gs}, \citenamefont {Gross}, \citenamefont {Huebl},\ and\ \citenamefont {Althammer}}]{hanle2}%
  \BibitemOpen
  \bibfield  {author} {\bibinfo {author} {\bibfnamefont {T.}~\bibnamefont {Wimmer}}, \bibinfo {author} {\bibfnamefont {A.}~\bibnamefont {Kamra}}, \bibinfo {author} {\bibfnamefont {J.}~\bibnamefont {G{\"u}ckelhorn}}, \bibinfo {author} {\bibfnamefont {M.}~\bibnamefont {Opel}}, \bibinfo {author} {\bibfnamefont {S.}~\bibnamefont {Gepr{\"a}gs}}, \bibinfo {author} {\bibfnamefont {R.}~\bibnamefont {Gross}}, \bibinfo {author} {\bibfnamefont {H.}~\bibnamefont {Huebl}},\ and\ \bibinfo {author} {\bibfnamefont {M.}~\bibnamefont {Althammer}},\ }\href@noop {} {\bibfield  {journal} {\bibinfo  {journal} {Observation of antiferromagnetic magnon pseudospin dynamics and the Hanle effect, Physical Review Letters}\ }\textbf {\bibinfo {volume} {125}},\ \bibinfo {pages} {247204} (\bibinfo {year} {2020})}\BibitemShut {NoStop}%
\end{thebibliography}%
\end{document}